\documentclass[aps,pre,reprint,superscriptaddress]{revtex4-2}

\usepackage{graphicx}
\usepackage{amssymb,amsmath}
\usepackage{bm,color}
\usepackage{multirow}

\begin{document}

\title{Control of phase ordering and elastic properties in phase field crystals through three-point direct correlation}

\author{Zi-Le Wang}
\affiliation{Technology and Engineering Center for Space Utilization, Chinese Academy of Sciences, Beijing 100094, China}
\affiliation{State Key Laboratory of Low Dimensional Quantum Physics, Department of Physics, Tsinghua University, Beijing 100084, China}
\author{Zhirong Liu}
\email{LiuZhiRong@pku.edu.cn}
\affiliation{College of Chemistry and Molecular Engineering, Peking University, Beijing 100871, China}
\author{Wenhui Duan}
\email{duanw@tsinghua.edu.cn}
\affiliation{State Key Laboratory of Low Dimensional Quantum Physics, Department of Physics, Tsinghua University, Beijing 100084, China}
\affiliation{Institute for Advanced Study, Tsinghua University, Beijing 100084, China}
\affiliation{Collaborative Innovation Center of Quantum Matter, Tsinghua University, Beijing 100084, China}
\affiliation{Frontier Science Center for Quantum Information, Beijing 100084, China}
\author{Zhi-Feng Huang}
\email{huang@wayne.edu}
\affiliation{Department of Physics and Astronomy, Wayne State University, Detroit, Michigan 48201, USA}

\date{\today}

\begin{abstract}

Effects of three-point direct correlation on properties of the phase field crystal (PFC) modeling are examined, for the control of various ordered and disordered phases and their coexistence in both three-dimensional and two-dimensional systems. Such effects are manifested via the corresponding gradient nonlinearity in the PFC free energy functional that is derived from classical density functional theory. Their significant impacts on the stability regimes of ordered phases, phase diagrams, and elastic properties of the system, as compared to those of the original PFC model, are revealed through systematic analyses and simulations. The nontrivial contribution from three-point direct correlation leads to the variation of the critical point of order-disorder transition to which all the phase boundaries in the temperature-density phase diagram converge. It also enables the variation and control of system elastic constants over a substantial range as needed in modeling different types of materials with the same crystalline structure but different elastic properties. The capability of this PFC approach in modeling both solid and soft matter systems is further demonstrated through the effect of three-point correlation on controlling the vapor-liquid-solid coexistence and transitions for body-centered cubic (bcc) phase and on achieving the liquid-stripe or liquid-lamellar phase coexistence. All these provide a valuable and efficient method for the study of structural ordering and evolution in various types of material systems.

\end{abstract}

\maketitle

\section{Introduction}
\label{sec: introduction}

The phase field crystal (PFC) method, an effective density-field approach for modeling atomistic details of material systems on diffusive time scales \cite{prl245701,pre51605,prb064107,huang08,prl35501}, has been widely applied to the study of various structural and dynamical phenomena in a broad range of areas, such as solidification \cite{prb064107,pre051404,pre031603,prm095603}, elastic and plastic deformation of materials \cite{prb184107,pre046107,prb014103,prb144112,pre013302,ZhouCarbon19}, crystal growth \cite{huang08,apa385,pre012405,pccp21858}, dislocation dynamics \cite{pre031609,prb224114,prl015502,prb054113,prb014107,SalvalaglioPRL21}, grain boundary structures and dynamics \cite{prb024110,prb035414,prl255501_2017,acta160,prm053804}, ferromagnetics and ferroelectrics \cite{prb184109}, quasicrystals \cite{prl255501_2014,prl075501}, heterostructures and stacked multilayers of two-dimensional (2D) materials \cite{prb165412,prm034004}, among many others. The PFC models can be connected to or derived from classical density functional theory (cDFT) through the expansion of direct correlation functions \cite{prb064107,pre051404,prl045702,prb180102}. The original PFC model contains only two-point direct correlation where crystal structures and ordered patterns are controlled by a single microscopic lattice length scale. This limited its capacity in simulating complex structures. Some works have been conducted to overcome this shortcoming. For example, the multi-mode PFC models have been proposed to describe a variety of complex structures such as honeycomb, kagome, rectangular, oblique, and other hybrid phases or chiral and achiral superlattices in 2D \cite{prl35501,MkhontaPRL16} and face-centered cubic (fcc) phase in three dimensions (3D) \cite{pre061601}. The XPFC model \cite{prl045702,pre031601} has the ability to stabilize the square phase in 2D and fcc, hexagonal close packed (hcp), or simple cubic structures in 3D. Compared to the original single-mode PFC model, the main difference in the multimode PFC or XPFC models is that there are more than one peak in the Fourier component of the two-point direct correlation function to represent multiple lattice length scales.

From a theoretical point of view, it would be necessary to consider multi-point direct correlations for the completeness of the PFC free energy functional. More importantly, the incorporation of multi-point correlations can enable the study of broader material systems and enrich the properties of the PFC modeling. Limited attempts have been made to explore the influence of multi-point correlations in PFC \cite{prb035447,prl155501,prm060801,prb180102,prm103802}. For example, Seymour and Provatas \cite{prb035447} approximated the three-point direct correlation function as the product of two-point direct correlations to model structures with a specific bond angle, such as triangular, honeycomb, or square, and Kocher and Provatas \cite{prl155501} used both three- and four-point direct correlations to model vapor-liquid-solid transitions. Alternatively, Alster {\it et al.} \cite{prm060801} expanded the three-point correlation function in terms of Legendre polynomials in Fourier space and constructed various crystalline phases (including the $ABX_3$ perovskite structure) from their PFC model. Recently, we developed a general PFC formulation to incorporate any multi-point direct correlations satisfying the condition of rotational invariance \cite{prb180102}, from which effects of bond-angle dependency and adjustment can be achieved through the four-point correlation, and a variety of 2D and 3D crystal structures, such as 3D diamond cubic phase and 2D rhombic or 3D simple monoclinic structure with tunable bond angles, can be stabilized. Based on this approach, an efficient PFC modeling for 2D vapor-liquid-solid coexistence and transitions was subsequently developed by considering the three-point direct correlation \cite{prm103802}.

Despite these recent progresses, the influence of multi-point correlations in PFC and the corresponding properties are still not well understood. For example, it is unclear how the phase diagram and the relative stability and coexistence between different ordered and disordered phases would change with the incorporation of multi-point correlations. In this study, we systematically investigate the influence of three-point direct correlation on the single-component PFC phase diagram, particularly the conditions of stability and coexistence among various phases. Our results reveal that the ordered vs disordered stability regime and the relative stability between ordered phases can be controlled by the three-point direct correlation (Sec.~\ref{sec:stability analysis}). In addition, contributions from the three-point correlation can notably impact the phase diagrams as compared to the original PFC model. When the effect of three-point correlation is incorporated, the critical point in the $\epsilon$ (temperature parameter) vs $\bar{n}$ (average density) phase diagram, at which all the phase boundaries intersect and the phase coexistence terminates, becomes variable (Sec.~\ref{sec:Phase_diagram}). The effect on system elastic properties is also revealed, showing as the ability to vary the elastic constants in the model over a considerably broad range for describing different materials of same crystalline symmetry (Sec.~\ref{sec:elasticity}). Other examples showing the control capability of three-point correlation include the vapor-liquid-solid transitions and coexistence for 3D body-centered cubic (bcc) phase (Sec.~\ref{sec:VLS_bcc}) and interestingly, the realization of liquid-stripe or liquid-lamellar coexistence (Sec.~\ref{sec:lid_str_coex}), both of which are verified by dynamical simulations of the full PFC model. These further demonstrate the applicability and efficiency of our PFC modeling for both solid and soft matter systems.

\section{Model}
\label{sec:model}

The rescaled free energy functional in the original PFC model is written as \cite{prl245701,pre51605,huang08}
\begin{equation}
\mathcal{F}[n]
=\int{d\mathbf{r}\left\{ \frac{1}{2}n \left[
-\epsilon+\left(\nabla^2+1\right)^2 \right]n - \frac{g}{3}n^3
+\frac{1}{4}n^4 \right\}},
\label{OriPFC}
\end{equation}
where $n(\mathbf{r},t)$ is an order parameter field representing atomic number density variation, $g$ is a phenomenological model parameter, and $\epsilon$ is a temperature parameter with larger (smaller) $\epsilon$ value corresponding to lower (higher) temperature. In the associated phase diagram the critical point is located at $\epsilon=0$ for the transition between disordered (liquid) and ordered phases.

This original PFC model can be derived from cDFT with two-point direct correlation, the effect of which yields the gradient terms of the PFC free energy functional in Eq.~(\ref{OriPFC}) \cite{prb064107}. As detailed in Ref.~\cite{prb180102}, a more general PFC-type density-field formulation can be developed to incorporate any orders of multi-point direct correlations that are rotationally invariant. Here we give a brief description to show how it leads to the free energy functional used in this study integrating the effect of three-point correlation.

Generally, the free energy functional in cDFT is expanded by \cite{PR351,prb180102}
\begin{eqnarray}
&& \frac{\Delta\mathcal{F}[n]}{k_BT}=
    \rho_0 \int d\mathbf{r} (1+n) \ln (1+n) - \sum_{m} \frac{\rho_0^m}{m!} \nonumber\\
&& \times \int \prod_{j=1}^m d\mathbf{r}_j
  C^{(m)}(\mathbf{r}_1, \mathbf{r}_2, ..., \mathbf{r}_m)
  n(\mathbf{r}_1) n(\mathbf{r}_2) \cdots n(\mathbf{r}_m), \label{eq:F}
\end{eqnarray}
where $\Delta\mathcal{F}=\mathcal{F}-\mathcal{F}_0$ with $\mathcal{F}_0$ the free energy at the reference-state density $\rho_0$, the density variation field $n=(\rho-\rho_0)/\rho_0$ with $\rho$ the atomic number density, and $C^{(m)}(\mathbf{r}_1, \mathbf{r}_2, ..., \mathbf{r}_m)$ is the $m$-point direct correlation function at the reference state. $C^{(m)}$ is translationally invariant and can be rewritten as $C^{(m)}(\mathbf{r}_1, \mathbf{r}_2, ..., \mathbf{r}_m) = C^{(m)}(\mathbf{r}_1-\mathbf{r}_2, \mathbf{r}_1-\mathbf{r}_3, ..., \mathbf{r}_1-\mathbf{r}_m)$ without loss of generality. Substituting its Fourier transform
\begin{eqnarray}
  && C^{(m)}(\mathbf{r}_1, \mathbf{r}_2, ..., \mathbf{r}_m) \nonumber\\
  && = \frac{1}{(2\pi)^{(m-1)d}} \int d\mathbf{q}_1 d\mathbf{q}_2 ...  d\mathbf{q}_{m-1} \hat{C}^{(m)} (\mathbf{q}_1, \mathbf{q}_2, ..., \mathbf{q}_{m-1}) \nonumber\\
  && \times e^{i \mathbf{q}_1 \cdot (\mathbf{r}_1-\mathbf{r}_2)} e^{i \mathbf{q}_2 \cdot (\mathbf{r}_1-\mathbf{r}_3)} ... e^{i \mathbf{q}_{m-1} \cdot (\mathbf{r}_1-\mathbf{r}_m)},
\end{eqnarray}
into Eq.~(\ref{eq:F}) yields
\begin{eqnarray}
  && \int \prod_{j=1}^m d\mathbf{r}_j
  C^{(m)}(\mathbf{r}_1, \mathbf{r}_2, ..., \mathbf{r}_m)
  n(\mathbf{r}_1) n(\mathbf{r}_2) \cdots n(\mathbf{r}_m) \nonumber\\
  && = \frac{1}{(2\pi)^{(m-1)d}}\int d\mathbf{r} n(\mathbf{r}) \int d\mathbf{q}_1 d\mathbf{q}_2 ...  d\mathbf{q}_{m-1}
  e^{i \mathbf{q}_1 \cdot \mathbf{r}}
  e^{i \mathbf{q}_2 \cdot \mathbf{r}} ... \nonumber\\
  && e^{i \mathbf{q}_{m-1} \cdot \mathbf{r}}
  \hat{C}^{(m)}(\mathbf{q}_1, \mathbf{q}_2, ..., \mathbf{q}_{m-1})
  \hat{n}(\mathbf{q}_1) \hat{n}(\mathbf{q}_2) ... \hat{n}(\mathbf{q}_{m-1}),
  \label{eq:C_int}
\end{eqnarray}
where $\hat{n}(\mathbf{q})$ is the Fourier transform of $n(\mathbf{r})$ and $d$ is the system dimensionality. The Fourier component $\hat{C}^{(m)}(\mathbf{q}_1, \mathbf{q}_2, ..., \mathbf{q}_{m-1})$ can be expanded as a power series of wave vector $\mathbf{q}_i$, the tensor form of which is given by
\begin{eqnarray}
&&\hat{C}^{(m)}(\mathbf{q}_1, \mathbf{q}_2, ..., \mathbf{q}_{m-1})
= \sum_{K=0}^\infty ~\sum_{i_1,i_2,...,i_K =1}^{m-1} \nonumber\\
&& \sum_{\alpha_{i_1},\alpha_{i_2},...,\alpha_{i_K} =x,y,z}
C_{i_1\alpha_{i_1} i_2\alpha_{i_2} ... i_K\alpha_{i_K}} T_{i_1\alpha_{i_1}  i_2\alpha_{i_2} ... i_K\alpha_{i_K}}^{(K)},
\end{eqnarray}
where $T_{i_1\alpha_{i_1} i_2\alpha_{i_2} ... i_K\alpha_{i_K}}^{(K)} = q_{i_1\alpha_{i_1}} q_{i_2\alpha_{i_2}} \cdots q_{i_K\alpha_{i_K}}$ are components of a rank-$K$ tensor $\mathbf{T}^{(K)}$. Both ${C}^{(m)}$ and $\hat{C}^{(m)}$ are rotationally invariant, and thus $\mathbf{T}^{(K)}$ should be a 2D or 3D isotropic Cartesian tensor which can be written as linear combinations of products of $\mathbf{q}_i \cdot \mathbf{q}_j$ and $(\mathbf{q}_{k} \times \mathbf{q}_{l}) \cdot \mathbf{q}_{p}$, leading to \cite{prb180102}
\begin{eqnarray}
  &&\hat{C}^{(m)}(\mathbf{q}_1, \mathbf{q}_2, ..., \mathbf{q}_{m-1}) \nonumber\\
  &&= \sum_{\mu_{11},\mu_{12},...=0}^{\infty} \hat{C}^{(m)}_{\mu_{11}\mu_{12}...}
  \prod_{i,j=1}^{m-1} \left (\mathbf{q}_i \cdot \mathbf{q}_j \right )^{\mu_{ij}} \label{eq:C_expan}\\
  &&+ \sum_{k,l,p=1}^{m-1} \sum_{\nu_{11},...=0}^{\infty} \hat{C}^{(m)}_{\nu_{11}...klp}
  \prod_{i,j=1}^{m-1} \left (\mathbf{q}_i \cdot \mathbf{q}_j \right )^{\nu_{ij}}
  \left [ \left (\mathbf{q}_{k} \times \mathbf{q}_{l} \right ) \cdot \mathbf{q}_{p} \right ]. \nonumber
\end{eqnarray}
Substituting it into Eq.~(\ref{eq:C_int}) and making use of some properties of Fourier transform [e.g., $\int d\mathbf{q} \exp(i \mathbf{q} \cdot \mathbf{r}) \mathbf{q}^k \hat{n}(\mathbf{q}) = (-i)^k {\bm \nabla}^k n(\mathbf{r})$], the corresponding PFC free energy functional with contributions from multi-point direct correlations can be obtained. Keeping the expansion up to four-point correlations and truncating at $\mathcal{O}(q^4)$, the resulting free energy functional is give by
\begin{eqnarray}
 && \mathcal{F}[n] =-\int B_0 n(\mathbf{r})d\mathbf{r} \nonumber\\
 && -\frac{1}{2} \int n(\mathbf{r})\left(C_0+C_2\nabla^2 + C_4\nabla^4 \right)
 n(\mathbf{r}) d\mathbf{r} \nonumber\\
 && -\frac{1}{3!} \int \Biglb\{ D_0 n^3(\mathbf{r}) + D_{11}n^2(\mathbf{r})\nabla^2 n(\mathbf{r}) \nonumber\\
 && \qquad\quad +D_{1111}n^2(\mathbf{r})\nabla^4 n(\mathbf{r})
 +D_{1122}n(\mathbf{r})\left[\nabla^2 n(\mathbf{r})\right]^2 \Bigrb\} d\mathbf{r} \nonumber\\
 && -\frac{1}{4!} \int \Biglb\{ E_0n^4(\mathbf{r})
 +E_{11}n^3(\mathbf{r})\nabla^2 n(\mathbf{r})
 +E_{1111}n^3(\mathbf{r})\nabla^4 n(\mathbf{r}) \nonumber\\
 && \qquad\quad +E_{1122}n^2(\mathbf{r})\left[\nabla^2 n(\mathbf{r})\right]^2
 +E_{44}n^2(\mathbf{r})\nabla^4 n^2(\mathbf{r})
 \Bigrb\} d\mathbf{r}, \nonumber \\
 \label{Functional_complete}
\end{eqnarray}
where the $C$ parameters (i.e., $C_0$, $C_2$, and $C_4$) are proportional to the expansion coefficient of two-point direct correlation $\hat{C}^{(2)}$ in Fourier space, and the $D$ and $E$ terms represent the contributions from three- and four-point direct correlations, respectively, with the corresponding parameters dependent on the expansion coefficients of $\hat{C}^{(3)}$ and $\hat{C}^{(4)}$ given in Eq.~(\ref{eq:C_expan}). Here the first term of Eq.~(\ref{eq:F}) [i.e., the ideal-gas entropy contribution $(1+n)\ln(1+n)$] has been expanded as a power series of $n$, as in the original PFC, and been merged into the $B_0$, $C_0$, $D_0$, and $E_0$ terms.

When keeping only the $D_0$ and $E_0$ terms as well as the $C$ terms from two-point correlation in Eq.~(\ref{Functional_complete}), we can recover the original PFC model after rescaling, i.e., Eq.~(\ref{OriPFC}), with $C_0=\epsilon-1$, $C_2=-2$, $C_4=-1$, $D_0=2g$, and $E_0=-6$. Therefore, the first nontrivial contribution from three-point direct correlation is the nonlinear gradient term with coefficient $D_{11}$, the effect of which is the focus of this study. In the following we use a reduced form of Eq.~(\ref{Functional_complete}) as the free-energy functional in our PFC approach, i.e.,
\begin{eqnarray}
 \mathcal{F}[n] = && -\int B_0 n d\mathbf{r} \nonumber\\
 && -\frac{1}{2} \int n \left(C_0+C_2\nabla^2 + C_4\nabla^4 + C_6\nabla^6 \right) n d\mathbf{r} \nonumber\\
 && -\frac{1}{6} \int \left( D_0 n^3 + D_{11}n^2\nabla^2 n
 \right ) d\mathbf{r} \nonumber\\
 && -\frac{1}{24} \int \left [ E_0n^4
 +E_{1122}n^2\left(\nabla^2 n\right)^2 \right ] d\mathbf{r},
 \label{Functional}
\end{eqnarray}
which has been shown in our recent work to well produce the vapor-liquid-solid coexistence and transitions for 2D triangular phase and some realistic temperature- and pressure-related material properties \cite{prm103802}. In this model $E_0<0$ and $E_{1122}\leq 0$ are required for convergence. The $E_{1122}$ term, another gradient nonlinearity originated from the contribution of four-point correlation, is used to prevent the divergence of ordered-phase free energy when the effect of three-point correlation, i.e., the $D_{11}$ term, is present. An additional $C_6$ term is introduced in Eq.~(\ref{Functional}), corresponding to the expansion of two-point correlation $\hat{C}^{(2)}$ at $\mathcal{O}(q^6)$. It is used to effectively suppress the contributions from high-order crystalline modes when $D_{11}$ and $E_{1122}$ terms are incorporated, as demonstrated in Ref.~\cite{prm103802}.

The dynamical evolution of the system is governed by the conserved PFC equation
\begin{equation}
\frac{\partial n}{\partial t}=\nabla^2 \frac{\delta\mathcal{F}[n]}{\delta n}.
\label{DynamEq}
\end{equation}
From the free energy functional in Eq.~(\ref{Functional}), we get
\begin{eqnarray}
  \frac{\partial n}{\partial t} &&= \nabla^2 \Biglb \{ - \left (C_0 + C_2 \nabla^2 + C_4 \nabla^4 + C_6 \nabla^6 \right ) n \nonumber\\
  && -\frac{1}{2} D_0 n^2 - \frac{1}{6} D_{11} \left ( 2n\nabla^2 n + \nabla^2 n^2 \right ) \nonumber\\
  && -\frac{1}{6} E_0 n^3 -\frac{1}{12} E_{1122}
     \left [ n (\nabla^2 n)^2 + \nabla^2 (n^2 \nabla^2 n) \right ] \Bigrb \}, \label{EqPFC}
\end{eqnarray}
which is used for our full-mode PFC simulations.

\section{Analyses and Results}

\subsection{One-mode approximation}

To conduct stability analysis, one-mode approximation is first used for the density variation field $n(\mathbf{r})$ of  periodic ordered phases, i.e.,
\begin{equation}
  n(\mathbf{r})=\bar{n} +A\sum_{\mathbf{q}}\left(e^{i\mathbf{q}\cdot\mathbf{r}}+\textrm{c.c.}\right), \label{n_expansion}
\end{equation}
where $\bar{n}$ is the average density variation, $A$ is the amplitude, $\mathbf{q}$ represents the basic wave vector of the periodic structure, and ``c.c.''~refers to the complex conjugate. For a 2D stripe or 3D lamellar phase with $\mathbf{q}=q(1,0,0)$, from Eq.~(\ref{Functional}) the corresponding free energy density is written as
\begin{eqnarray}
  && f_{\rm stripe}\left(q,A;\bar n \right)
  =-B_0\bar n-\frac{1}{2}C_0 \bar n^2-\frac{1}{6}D_0 \bar n^3 - \frac{1}{24} E_0 \bar n^4 \nonumber \\
  && -\left[\left(C_0-C_2q^2+C_4q^4-C_6q^6\right)+\frac{1}{3}\left(3D_0-2D_{11}q^2\right)\bar n
    \right. \nonumber\\
  && \left. +\frac{1}{12}\left(6E_0+E_{1122}q^4\right)\bar n^2\right]A^2
  -\frac{1}{4}\left(E_0+E_{1122}q^4\right)A^4. \nonumber\\
  \label{F_Stripes}
\end{eqnarray}

For 2D triangular or 3D rod phase with basic wave vectors $\mathbf{q}=q(1,0,0)$ and  $q(1/2,\pm \sqrt{3}/2,0)$, the free energy density is obtained as
\begin{eqnarray}
&& f_{\rm tri}\left(q,A;\bar n \right)
  =-B_0\bar n-\frac{1}{2}C_0 \bar n^2-\frac{1}{6}D_0 \bar n^3 - \frac{1}{24} E_0 \bar n^4 \nonumber \\
  && -3\left[\left(C_0-C_2q^2+C_4q^4-C_6q^6\right)+\frac{1}{3}\left(3D_0-2D_{11}q^2\right)\bar{n} \right. \nonumber\\
  && \quad\quad \left.+\frac{1}{12}\left(6E_0+E_{1122}q^4\right)\bar n^2\right]A^2 \nonumber\\
  && -\left[2\left(D_0-D_{11}q^2\right)+\left(2E_0+E_{1122}q^4\right)\bar n\right]A^3 \nonumber \\
  && -\frac{15}{4}\left(E_0+E_{1122}q^4\right)A^4. \label{F_Rods}
\end{eqnarray}
The inverse of triangular structure is honeycomb, which has the same basic wave vectors but with opposite sign of amplitude $A$. Thus, if changing $A$ to $-A$ in Eq.~(\ref{n_expansion}) to maintain positive amplitude, the free energy density becomes
\begin{eqnarray}
&& f_{\rm hon}\left(q,A;\bar n \right)=-B_0\bar n-\frac{1}{2}C_0 \bar n^2-\frac{1}{6}D_0 \bar n^3 - \frac{1}{24} E_0 \bar n^4 \nonumber \\
  && -3\left[\left(C_0-C_2q^2+C_4q^4-C_6q^6\right)+\frac{1}{3}\left(3D_0-2D_{11}q^2\right)\bar n \right.  \nonumber\\
  && \quad\quad \left.+\frac{1}{12}\left(6E_0+E_{1122}q^4\right)\bar n^2\right]A^2 \nonumber\\
  && +\left[2\left(D_0-D_{11}q^2\right)+\left(2E_0+E_{1122}q^4\right)\bar n\right]A^3 \nonumber \\
  && -\frac{15}{4}\left(E_0+E_{1122}q^4\right)A^4. \label{F_graphene}
\end{eqnarray}
Note that the only difference between Eq.~(\ref{F_graphene}) and Eq.~(\ref{F_Rods}) is the sign of the $A^3$ term.

\begin{figure}[htb]
  \centerline{\includegraphics[width=0.5\textwidth]{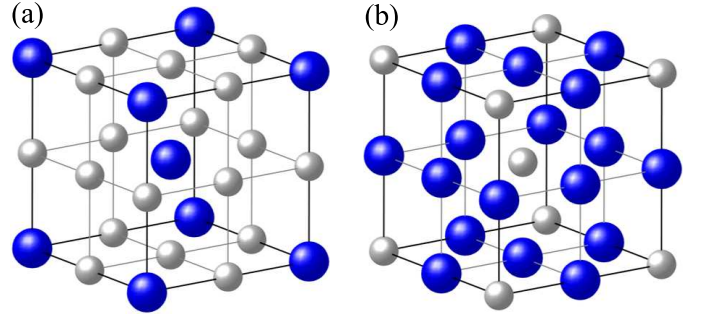}}
  \caption{The real-space lattice structures of (a) bcc and (b) mbcc phases. The blue spheres represent positions of atoms with maximum density $n$, while the gray spheres represent the positions of density minimum.}
  \label{fig:Fig1}
\end{figure}

For 3D bcc phase, the one-mode basic wave vectors $\mathbf{q}=q(1,\pm 1,0)/\sqrt{2}$, $q(0,1,\pm 1)/\sqrt{2}$, and $q(\pm 1,0,1)/\sqrt{2}$; thus its free energy density is given by
\begin{eqnarray}
&& f_{\rm bcc}\left(q,A;\bar n \right)
  =-B_0\bar n-\frac{1}{2}C_0 \bar n^2-\frac{1}{6}D_0 \bar n^3 - \frac{1}{24} E_0 \bar n^4 \nonumber \\
  && -6\left[\left(C_0-C_2q^2+C_4q^4-C_6q^6\right)+\frac{1}{3}\left(3D_0-2D_{11}q^2\right)\bar{n} \right. \nonumber\\
  && \quad\quad \left.+\frac{1}{12}\left(6E_0+E_{1122}q^4\right)\bar n^2\right]A^2 \nonumber\\
  && -4\left[2\left(D_0-D_{11}q^2\right)+\left(2E_0+E_{1122}q^4\right)\bar n\right]A^3 \nonumber \\
  && -\frac{45}{2}\left(E_0+E_{1122}q^4\right)A^4.
\label{F_bcc}
\end{eqnarray}
For the inverse of bcc structure, the maximum and minimum locations of the atomic density $n$ are expected to be reversed as compared to bcc (see Fig.~\ref{fig:Fig1}), similar to the relation between honeycomb and triangular phases described above. Thus this inverse phase should have the same basic wave vectors $\mathbf{q}$ as bcc but a negative value of $A$. Similarly, replacing $A$ in the expansion of Eq.~(\ref{n_expansion}) by $-A$ and substituting into Eq.~(\ref{Functional}), we obtain the following expression of free energy density for this inverse bcc phase which is now named mbcc (where ``m" refers to ``minus"):
\begin{eqnarray}
&& f_{\rm mbcc}\left(q,A;\bar n \right)
  =-B_0\bar n-\frac{1}{2}C_0 \bar n^2-\frac{1}{6}D_0 \bar n^3 - \frac{1}{24} E_0 \bar n^4 \nonumber \\
  && -6\left[\left(C_0-C_2q^2+C_4q^4-C_6q^6\right)+\frac{1}{3}\left(3D_0-2D_{11}q^2\right)\bar{n} \right. \nonumber\\
  && \quad\quad \left.+\frac{1}{12}\left(6E_0+E_{1122}q^4\right)\bar n^2\right]A^2 \nonumber\\
  && +4\left[2\left(D_0-D_{11}q^2\right)+\left(2E_0+E_{1122}q^4\right)\bar n\right]A^3 \nonumber \\
  && -\frac{45}{2}\left(E_0+E_{1122}q^4\right)A^4.
\label{F_mbcc}
\end{eqnarray}
Again, Eqs.~(\ref{F_bcc}) and (\ref{F_mbcc}) differ only in the sign of the $A^3$ term, as expected from their reverse of structure.

The free energy density of uniform or homogeneous phase, either liquid or vapor, is simply given by
\begin{equation}
f_u(\bar{n})=-B_0\bar{n}-\frac{1}{2}C_0\bar{n}^2-\frac{1}{6}D_0\bar{n}^3-\frac{1}{24}E_0\bar{n}^4.
\label{fl}
\end{equation}
 The vapor-liquid coexistence can be identified by applying the common tangent rule on $f_u(\bar{n})$ in Eq.~(\ref{fl}), giving the values of coexistence density as
\begin{equation}
\bar n_{\rm coexist}=\left (-D_0\pm \sqrt{3D_{0}^{2}-6C_0E_0} \right ) /{E_0}.
\label{ncoexist}
\end{equation}
From $\partial^2f_u/\partial\bar{n}^2=0$ we can obtain the spinodal densities, i.e.,
\begin{equation}
\bar n_{\rm spinodal}=\left (-D_0\pm \sqrt{D_{0}^{2}-2C_0E_0} \right ) /{E_0}.
\label{nspinodal}
\end{equation}

\subsection{Stability analysis: Effect of three-point direct correlation}
\label{sec:stability analysis}

In order to examine how the $D_{11}$ term, which represents the effect of three-point direct correlation as described in Sec.~\ref{sec:model}, affects the stability regime of ordered phases, we conduct a stability analysis with respect to the supercooled or supersaturated uniform phase. Interestingly, the free energy densities given in Eqs.~(\ref{F_Stripes})--(\ref{F_mbcc}) show an intriguing property that either one of the amplitude expansion terms (i.e., $A^2$, $A^3$, or $A^4$ term) is proportional to a same factor for all the ordered phases under one-mode approximation. The linear stability is controlled by the $A^2$ term in Eqs.~(\ref{F_Stripes})--(\ref{F_mbcc}), all of which are proportional to the same factor
\begin{eqnarray}
\alpha_2(q,\bar n) = - && \left[\left(C_0-C_2q^2+C_4q^4\right)+\frac{1}{3}\left(3D_0-2D_{11}q^2\right)
  \bar n \right. \nonumber\\
&& \left. +\frac{1}{12}\left(6E_0+E_{1122}q^4\right)\bar n^2\right],
\label{Linear_termA2}
\end{eqnarray}
where we have set $C_6=0$ for simplicity, since nonzero $C_6$ would not qualitatively change the results as has been discussed in detail in Ref.~\cite{prm103802}. Briefly speaking, the introduction of the $C_6$ term is to enhance the stability of the phase behavior determined by one-mode analysis when solving the full PFC model via Eq.~(\ref{EqPFC}), such as the dynamic evolution of bcc and stripe or lamellar phases discussed below in Secs.~\ref{sec:VLS_bcc} and \ref{sec:lid_str_coex}. A nonzero $C_6$ could alter the degree of contributions from higher-order modes. The larger the $C_6$ value, the more effective suppression on the higher-order modes can be achieved. Results from the full model with large enough $C_6$ are very close to those of one-mode approximation, as verified numerically \cite{prm103802}. Thus the analytic stability results identified in this subsection with the absence of $C_6$, which are based on one-mode analysis, are used as the basis for the next-step full-model analysis and simulations given in Secs.~\ref{sec:VLS_bcc} and \ref{sec:lid_str_coex}.

The uniform phase (either liquid or vapor) is linearly unstable when subjected to infinitesimal fluctuations if $\alpha_2 < 0$, leading to the formation of an ordered phase. By solving $\partial \alpha_2/\partial q=0$ we can find the wave number $q$ for maximum instability, i.e.,
\begin{equation}
q=\left\{
\begin{array}{ll}
\sqrt{\frac{6C_2+4D_{11}\bar n}{12C_4+E_{1122}{\bar n}^2}}, & {\rm if~} 3C_2+2D_{11}\bar n <0,\\
0, & {\rm otherwise},
\end{array} \right.
\label{Linear_q2}
\end{equation}
given the condition $12C_4+E_{1122}{\bar n}^2<0$ for the convergence at large $q$ as seen in Eq.~(\ref{Linear_termA2}). When $3C_2+2D_{11}\bar n <0$, we then get
\begin{equation}
  \alpha_2(\bar n)= -\left(C_0+D_0\bar n+\frac{1}{2}E_0\bar n^2\right)
  +\frac{\left(3C_2+2D_{11}\bar n\right)^2}{3\left(12C_4+E_{1122}{\bar n}^2\right)}.
  \label{Linear_no_q}
\end{equation}
When $E_{1122}=0$ as assumed in this subsection (so that we could focus on the effect of $D_{11}$), $\alpha_2(\bar{n})$ in Eq.~(\ref{Linear_no_q}) remains a quadratic form of $\bar{n}$. The supersaturating or supercooling density for the occurrence of linear instability is obtained by solving $\alpha_2(\bar{n})=0$, which yields
\begin{equation}
\bar n_{\rm supercool} = \bar{n}_{\pm}
=\left ( -b_s \pm \sqrt{b_s^2-2a_sc_s} \right) /a_s,
\label{nsupercool}
\end{equation}
where $a_s=E_0-2D_{11}^2/(9C_4)$, $b_s=D_0-C_2D_{11}/(3C_4)$, and $c_s=C_0-C_2^2/(4C_4)$. We set $\bar{n}_+$ as the value of $\bar{n}_{\rm supercool}$ when the ``$+$'' sign is taken in Eq.~(\ref{nsupercool}), and set $\bar{n}_-$ as the value corresponding to the ``$-$'' sign.

In the original PFC model with $D_0=D_{11}=E_{1122}=0$, $q^2=C_2/(2C_4)$ according to Eq.~(\ref{Linear_q2}), independent of $\bar{n}$. There are two necessary conditions to stabilize ordered phases. The first one is $q^2>0$ which requires $C_2<0$ since $C_4<0$ is needed to prevent the divergence at large $q$ when $E_{1122}=0$. The second one is $\alpha_2(\bar{n})<0$. Since $E_0<0$ and hence $\alpha_2(\bar{n})$ is a convex parabola when $D_{11}=E_{1122}=0$, $\alpha_2(\bar{n})<0$ occurs between the two values of $\bar{n}_{\rm supercool}$ when $D_0^2-2E_0(C_0-C_2^2/(4C_4))>0$ based on Eq.~(\ref{nsupercool}). That is, we need $C_2<0$ and $C_0>C_2^2/(2C_4)$ for the stability of ordered phases with $\bar{n}$ chosen in between $\bar{n}_+$ and $\bar{n}_-$.

When $D_{11} \neq 0$, $3C_2+2D_{11}\bar{n}<0$ is required for $q^2>0$. We define $\bar{n}_0$ as the solution of $q^2=0$ for $\bar{n}$, as determined by Eq.~(\ref{Linear_q2}). Also, $\alpha_2(\bar{n})<0$ is needed for stabilizing the ordered phases as discussed above, leading to two situations for nonzero $D_{11}$. First, $\alpha_2(\bar{n})$ is a convex parabola if $a_s=E_0-2D_{11}^2/(9C_4)<0$, and thus $\alpha_2(\bar{n})<0$ occurs for $\bar{n}$ lying within $(\bar{n}_+, \bar{n}_-)$ when $b_s^2-2a_sc_s>0$. Secondly, $\alpha_2(\bar{n})$ is a concave parabola if $a_s=E_0-2D_{11}^2/(9C_4)>0$. When $b_s^2-2a_sc_s>0$, $\alpha_2(\bar{n})<0$ occurs in the region $(-\infty, \bar{n}_-)\cup(\bar{n}_+, +\infty)$. When $b_s^2-2a_sc_s<0$, $\alpha_2(\bar{n})$ is negative at every $\bar{n}$ value. The intersection between $q^2>0$ and $\alpha_2(\bar{n})<0$ determines the stability regime of ordered phases.

\begin{figure*}[htb]
  \centerline{\includegraphics[width=\textwidth]{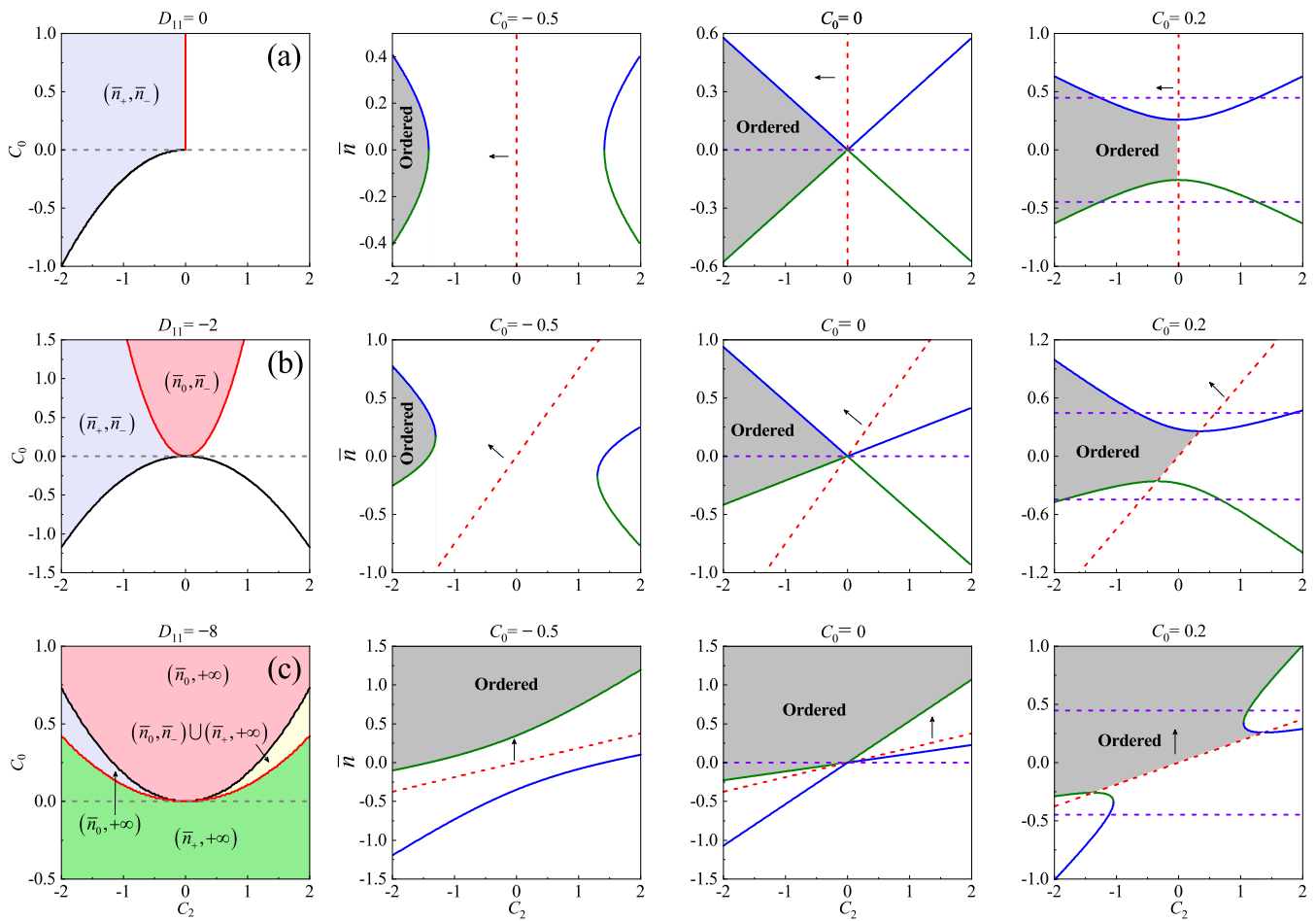}}
    \caption{Stability diagrams of ordered vs disordered phase obtained from linear instability analysis, at $C_6=D_0=E_{1122}=0$, $C_4=-1$, $E_0=-6$, and (a) $D_{11}=0$, (b) $D_{11}=-2$, and (c) $D_{11}=-8$. The first column shows the $C_0$-$C_2$ diagrams, with the colored areas corresponding to the stable regime of any ordered phases and white areas to that of the uniform phase. Different colored areas correspond to different linear stability regimes with the range of allowed $\bar{n}$ marked inside. The black and red curves correspond to $C_0=C_2^2/(4C_4)+[D_0-C_2D_{11}/(3C_4)]^2/[2E_0-4D_{11}^2/(9C_4)]$ and $C_0=3C_2D_0/(2D_{11})-9C_2^2E_0/(8D_{11}^2)$, respectively (see Appendix \ref{sec:Stable_region}). The horizontal dashed line indicates the value of $C_0$ at which the vapor-liquid phase separation takes place. All the other columns give the $\bar{n}$-$C_2$ stability diagrams, for different values of $C_0=-0.5$ (second column), $C_0=0$ (third column), and $C_0=0.2$ (fourth column). The linear stability regimes for the ordered phase are shown in shadow. The green curves refer to $\bar{n}_+$ and the blue ones represent $\bar{n}_-$, which are the values of $\bar{n}_{\rm supercool}$ calculated from Eq.~(\ref{nsupercool}). The purple dashed lines correspond to vapor-liquid coexistence densities obtained from Eq.~(\ref{ncoexist}), and the red dashed lines are for $3C_2+2D_{11}\bar{n}=0$. The arrows point to the region of $3C_2+2D_{11}\bar{n}<0$.}
\label{fig:Fig2}
\end{figure*}

\begin{table*}[t]
\centering
\caption{\label{tab:stab_reg}Stability regimes of ordered phase obtained from linear instability analysis, where $C_{01}=C_2^2/(4C_4)+b_s^2/(2a_s)$ and $C_{02}=3C_2D_0/(2D_{11})-9C_2^2E_0/(8D_{11}^2)$.}
\renewcommand\arraystretch{2}
\resizebox{\textwidth}{!}{
\begin{tabular}{|c|c|c|c|c|c|}
\hline
\multicolumn{2}{|c|}{\multirow{2}{*}{}}& \multicolumn{2}{c|}{$E_0-\frac{2D_{11}^2}{9C_4}<0$}&\multicolumn{2}{c|}{$E_0-\frac{2D_{11}^2}{9C_4}>0$}\\
\cline{3-6}

\multicolumn{2}{|c|}{} & $-\frac{\sqrt{9E_0C_4}}{2}<D_{11}<0$ & $0<D_{11}<\frac{\sqrt{9E_0C_4}}{2}$ & $D_{11}<-\frac{\sqrt{9E_0C_4}}{2}$ & $D_{11}>\frac{\sqrt{9E_0C_4}}{2}$ \\
\hline

\multirow{2}{*}{$C_2<\frac{2D_0D_{11}}{3E_0}$} & $C_{01}<C_0<C_{02}$ & $(\bar{n}_+, \bar{n}_-)$ & -
 & - & -  \\
\cline{2-6}

\multirow{2}{*}{} & $C_{02}<C_0<C_{01}$ & - & -  & $(\bar{n}_0, +\infty)$ &$(-\infty, \bar{n}_0)$ \\
\hline

\multirow{2}{*}{$C_2>\frac{2D_0D_{11}}{3E_0}$} & $C_{01}<C_0<C_{02}$ & - & $(\bar{n}_+, \bar{n}_-)$ & -
 & - \\
\cline{2-6}

\multirow{2}{*}{} & $C_{02}<C_0<C_{01}$ & - & - &$(\bar{n}_0, \bar{n}_-)\cup(\bar{n}_+, +\infty)$& $(-\infty, \bar{n}_-)\cup(\bar{n}_+, \bar{n}_0)$  \\
\hline

\multicolumn{2}{|c|}{$C_0>C_{02}$} & $(\bar{n}_0, \bar{n}_-)$ & $(\bar{n}_+, \bar{n}_0)$ & - & - \\
\hline

\multicolumn{2}{|c|}{$C_0<C_{02}$} & - & - & $(\bar{n}_+, +\infty)$ & $(-\infty, \bar{n}_-)$ \\
\hline

\multicolumn{2}{|c|}{$C_0>C_{01}$} & - & - & $(\bar{n}_0, +\infty)$ & $(-\infty, \bar{n}_0)$ \\
\hline

\end{tabular}
}
\end{table*}

These linear stability results are shown in Fig.~\ref{fig:Fig2} and summarized in Table \ref{tab:stab_reg}, with details of derivation given in Appendix \ref{sec:Stable_region}. There are two key curves (black and red, respectively) in each panel of the first column of Fig.~\ref{fig:Fig2} that separate the linear stability regimes for ordered and uniform phases, based on the analysis presented in Appendix \ref{sec:Stable_region}. Figure \ref{fig:Fig2}(a), i.e., the first row of Fig.~\ref{fig:Fig2}, gives the results of the original PFC model with $D_{11}=0$, for a range of $C_2$ values (noting that $C_2$ was restricted to a fixed value of $-2$ in almost all the previous studies of the original PFC). Since here $D_{11}=0$, $C_2$ should be negative to enable ordered phase according to Eq.~(\ref{Linear_q2}). The midpoint of two $\bar{n}_{\rm supercool}$ values (i.e., $\bar{n}_+$ and $\bar{n}_-$) coincides with that of $\bar{n}_{\rm coexist}$ for vapor-liquid coexistence, as can be seen in the last two columns of Fig.~\ref{fig:Fig2}(a). Therefore, when $D_{11}=0$ the stability regime of any solid or ordered phase is always found between vapor and liquid regimes, which results in a vapor-solid-liquid phase transition sequence \cite{prm103802}.

When $D_{11} \neq 0$ [Fig.~\ref{fig:Fig2}(b) and 2(c)], $C_2<0$ is no longer required for ordered phases. The midpoint of two $\bar{n}_{\rm supercool}$ values does not necessarily coincide with that of $\bar{n}_{\rm coexist}$ [see the last two columns of Fig.~\ref{fig:Fig2}(b) and 2(c)], and the areas of ordered-phase stability regime change. Importantly, if $D_{11}^2$ is larger than a critical value of $9C_4E_0/2$ (at which $a_s=E_0-2D_{11}^2/(9C_4)=0$), an ordered phase could be stabilized for any real value of $C_2$, as seen in Fig.~\ref{fig:Fig2}(c). In addition, the stability regime of the ordered phase with respect to the uniform phase (vapor or liquid) could be outside of the range confined by $(\bar{n}_+, \bar{n}_-)$, rather than lying between them as in the original PFC model. This makes it possible to choose parameters to obtain the usual sequence of vapor-liquid-solid transition as $\bar{n}$ increases [see the last column of Fig.~\ref{fig:Fig2}(c)].

\begin{figure*}[htb]
  \centerline{\includegraphics[width=\textwidth]{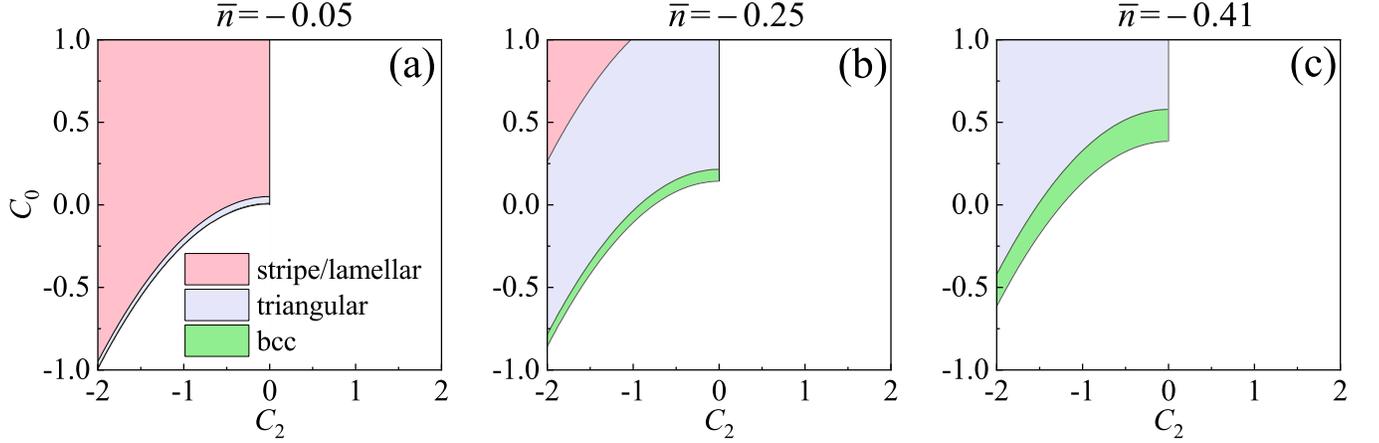}}
  \caption{Stability diagrams showing stability regimes for different ordered phases, with $D_{11}=0$ and $\bar{n}=-0.05$, $-0.25$, and $-0.41$ for (a), (b), and (c), respectively. The stability regimes are colored pink for stripe/lamellar phase, lavender for triangular, and green for bcc. Other parameters are set as $C_6=D_0=E_{1122}=0$, $C_4=-1$, and $E_0=-6$.}
  \label{fig:Fig3}
\end{figure*}

\begin{figure*}[htb]
  \centerline{\includegraphics[width=\textwidth]{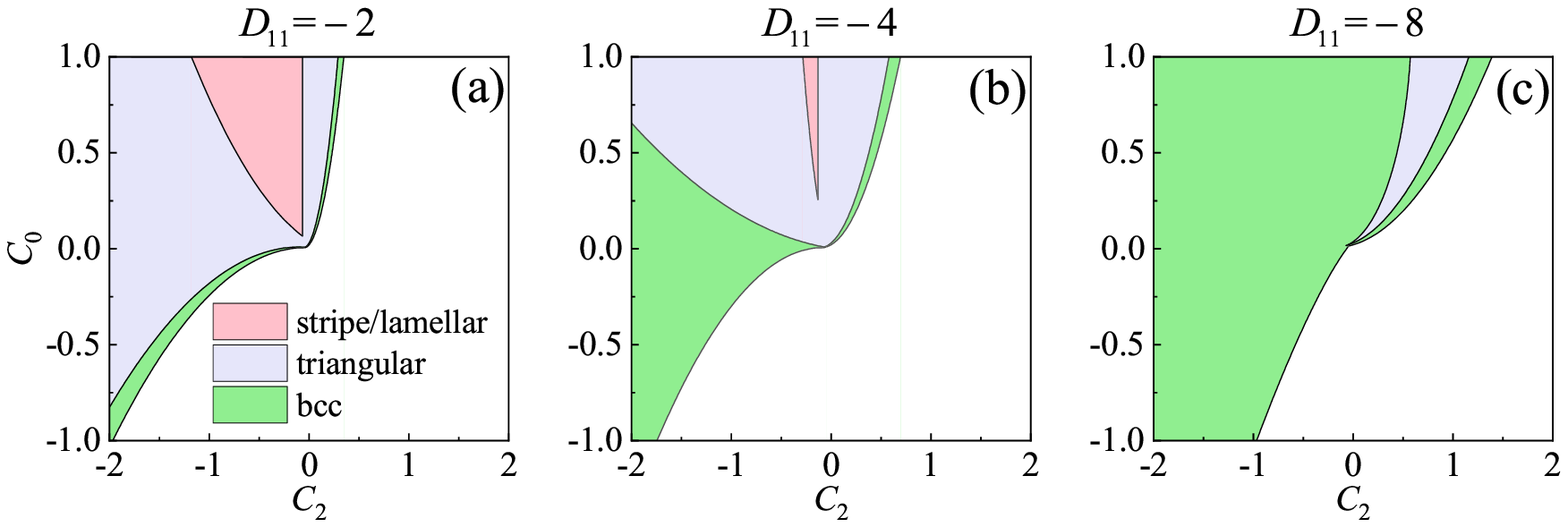}}
  \caption{Stability diagrams showing stability regimes for different ordered phases at a fixed value of $\bar{n}=-0.05$ and three different values of $D_{11}=-2$, $-4$, and $-8$ for (a), (b), and (c), respectively. The stability regimes are colored pink for stripe/lamellar phase, lavender for triangular, and green for bcc. Other parameters are set as $C_6=D_0=0$, $C_4=-1$, $E_0=-6$, and $E_{1122}=0$ for (a) and (b). In (c) $E_{1122}=-1$ is used to suppress the divergence of free energy densities of the ordered phases.
  }
  \label{fig:Fig4}
\end{figure*}

In the above analysis and results, only the linear stability of ordered vs disordered phase is identified, but not the relative stability between different individual ordered phases. To distinguish the phase stability among all the phases, we need to consider nonlinear contributions from all the terms in the free energy density of each ordered phase, i.e., $A^3$ and $A^4$ terms in Eqs.~(\ref{F_Stripes})--(\ref{F_mbcc}), in addition to the $A^2$ term used in linear stability analysis. The most stable phase at each point of the parameter space can then be determined through the minimum free energy density, which is used to obtain the full stability diagram. Some sample results of this full stability calculation are shown in Figs.~\ref{fig:Fig3} and \ref{fig:Fig4}. We first examine the case of $D_{11}=0$ (i.e., the original PFC model) for three negative values of $\bar{n}$. As seen in Figs.~\ref{fig:Fig3}(a)--\ref{fig:Fig3}(c), the stability regime for stripe or lamellar phase shrinks while those for bcc and triangular phases expand with the decrease of $\bar{n}$. Next, we incorporate the effect of nonzero $D_{11}$ in Fig.~\ref{fig:Fig4}, by choosing a fixed value of $\bar{n}=-0.05$ as an example to show the difference between various ordered-phase stability regimes as $D_{11}$ varies. If comparing Fig.~\ref{fig:Fig3}(a) for $D_{11}=0$ and Figs.~\ref{fig:Fig4}(a)--\ref{fig:Fig4}(c) for $D_{11}$ ranging from $-2$ to $-8$, it can be found that when the value of $D_{11}$ decreases, the range of stability for stripe/lamellar phase is reduced, accompanied by the expansion of the bcc stability regime. The regime for stripe/lamellar phase disappears completely at $D_{11}=-8$, with bcc phase occupying most of the stability regime of ordered phases [see Fig.~\ref{fig:Fig4}(c)]. This suggests that the $D_{11}$ term can be used to effectively control the relative stability between different ordered structures.

\subsection{$\epsilon$-$\bar{n}$ phase diagrams}
\label{sec:Phase_diagram}

\begin{figure*}
  \centerline{\includegraphics[width=1.0\textwidth]{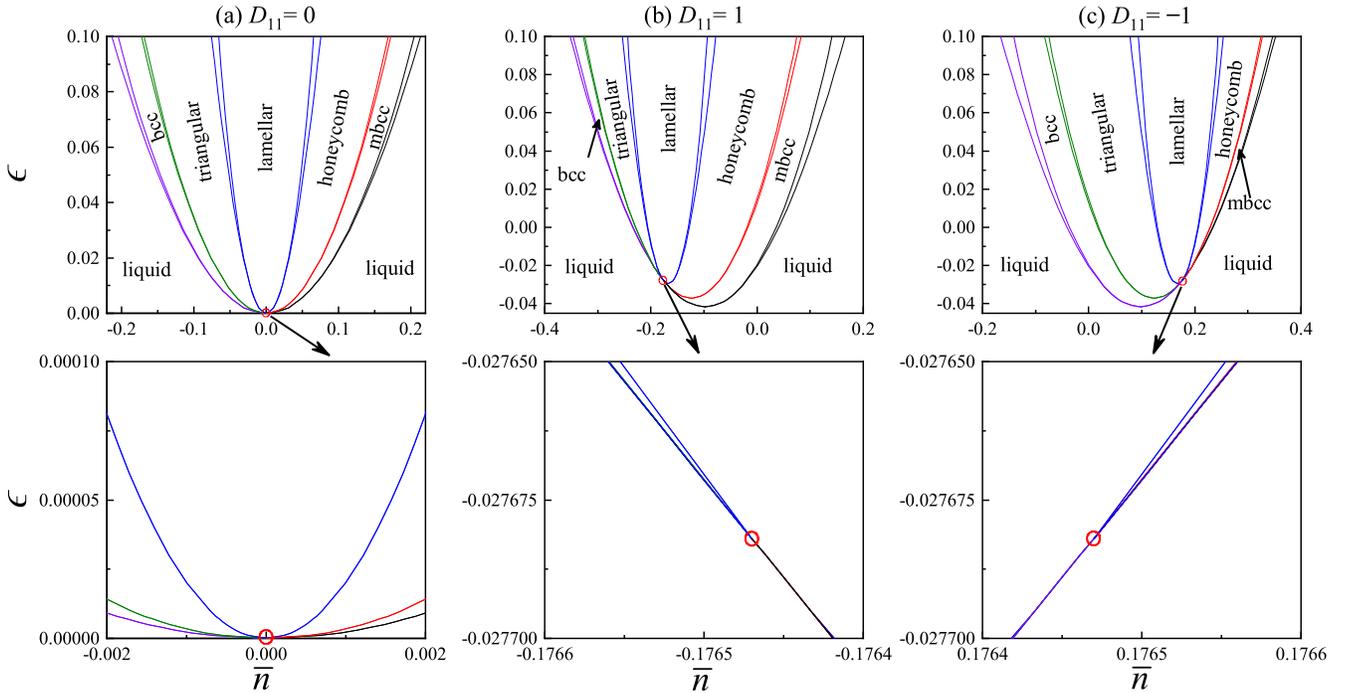}}
    \caption{Phase diagrams of 3D PFC calculated by one-mode approximation, for (a) $D_{11}=0$, (b) $D_{11}=1$, and (c) $D_{11}=-1$. Other parameters used are $B_0=C_6=D_0=E_{1122}=0$, $C_2=-2$, $C_4=-1$, and $E_0=-6$. Bottom panels show the enlarged regimes near the critical point (marked with red circle) in (a)--(c).}
\label{fig:Fig5}
\end{figure*}

The corresponding $\epsilon$-$\bar{n}$ (temperature-density) phase diagrams are presented in Fig.~\ref{fig:Fig5} (with $\epsilon=C_0+1$), showing the effect of the $D_{11}$ term on the stability and coexistence between different 3D phases including bcc, mbcc, lamellar, triangular rod, honeycomb rod, and liquid phases. The phase boundaries are determined by the common tangent construction based on the one-mode free energy densities in Eqs.~(\ref{F_Stripes})--(\ref{F_mbcc}).

It is interesting to note that there exists a special point in the phase diagram (marked as a red circle in Fig.~\ref{fig:Fig5}) at which all the phase boundaries coincide, i.e., all the ordered and disordered phases are indistinguishable and coexist at this point. This is a critical point where any two-phase coexistence terminates, with the same, unique location for all the phases. When $D_{11}=0$ it is located at $(\epsilon_c,\bar{n}_c)=(0,0)$ as known for the original PFC model [Fig.~\ref{fig:Fig5}(a)], while its location is adjustable via the variation of nonzero $D_{11}$ (i.e., via varying the contribution from three-point correlation), as seen by comparing Figs.~\ref{fig:Fig5}(b) and \ref{fig:Fig5}(c).
For the parameters used in Fig.~\ref{fig:Fig5} with $B_0=C_6=D_0=E_{1122}=0$, $C_2=-2$, $C_4=-1$, and $E_0=-6$, this critical point can be obtained analytically in the one-mode approximation, yielding
\begin{eqnarray}
&& \epsilon_c=1+\frac{27(D_{11}^2-12)}{(D_{11}^2-18)^2}, \label{eps_special_point}\\
&& \bar{n}_c=\frac{3 \left [ D_{11} \pm 9\sqrt{D_{11}^2/(D_{11}^2-18)^2} \right ]}{D_{11}^2-27},
\label{nbar_special_point}
\end{eqnarray}
with ``$+$'' for $D_{11} \geq 0$ and ``$-$'' for $D_{11} < 0$, which well agree with the numerical results given in Fig.~\ref{fig:Fig5} with high accuracy. The result of the critical point in the original PFC model, $\epsilon_c=\bar{n}_c=0$, can be recovered from Eqs.~(\ref{eps_special_point}) and (\ref{nbar_special_point}) by setting $D_{11}=0$. Details of derivation are provided in Appendix \ref{sec:Special_Point}, while some arguments for the corresponding governing conditions are explained below.

From the stability analysis given above in Sec.~\ref{sec:stability analysis}, the liquid-stripe or liquid-lamellar transition occurs at the supercooled point $\bar{n}_{\rm supercool}$ with the coefficient of the $A^2$ term $\alpha_2=0$. However, this is not the case for other ordered phases of which the free energy density contains an $A^3$ term [see Eqs.~(\ref{F_Rods})--(\ref{F_mbcc}) for hexagonal (triangular or honeycomb) rods, bcc, and mbcc phases]. The corresponding coefficient is proportional to
\begin{equation}
\alpha_3(q,\bar n) = 2\left(D_0-D_{11}q^2\right)+\left(2E_0+E_{1122}q^4\right)\bar n.
\label{Linear_termA3}
\end{equation}
The existence of an $A^3$ term can make an ordered phase at $\bar{n}_{\rm supercool}$ more stable than the liquid phase beyond the linear stability, since $A^3$ can take either positive or negative value to decrease the free energy density. To illustrate this, here we consider the case of $a_s=E_0-2D_{11}^2/(9C_4)<0$ as an example. In this case the linear stability regime of ordered phase is $(\bar{n}_+, \bar{n}_-)$ as described above. In all the free energy densities given in Eqs.~(\ref{F_Stripes})--(\ref{F_mbcc}), the $A^4$ term should be always positive to prevent the free-energy divergence. When $\bar{n}<\bar{n}_+$, the coefficient of $A^2$ term for all the ordered phases is positive, and thus the stripe or lamellar phase (without the $A^3$ term) is unstable with respect to liquid. The contribution from the $A^3$ term could be negative for hexagonal and bcc/mbcc phases within a certain range of $\bar{n}$ when $\bar{n}<\bar{n}_+$, such that these phases can be more stable than the liquid phase in this range. Therefore, the full stability range of these ordered phases with respective to liquid is broader than that determined by $\bar{n}_{\rm supercool}$. Usually this $\bar{n}$ range of stability is wider for the bcc/mbcc phase as compared to the hexagonal phase when $D_{11}=0$, leading to the liquid-bcc coexistence (instead of liquid-hexagonal coexistence) in 3D. The relative stability can be changed by nonzero $D_{11}$, as will be demonstrated below in Sec.~\ref{sec:lid_str_coex} where the liquid-lamellar or liquid-stripe coexistence is realized.

According to the above analysis, when $\alpha_3$ becomes zero under some conditions of e.g., $\bar{n}$ and $\epsilon$, the extra stabilization effect provided by the $A^3$ term no longer exists. Thus, to make the transition between all the ordered phases and the liquid phase occur at exactly the same point (i.e., controlled by $\bar{n}_{\rm supercool}$), we need the coefficients of $A^2$ and $A^3$ terms being equal to zero simultaneously, giving the following conditions in one-mode approximation
\begin{equation}
    \alpha_2(q_{\rm eq},\bar n) =0, \qquad
    \alpha_3(q_{\rm eq},\bar n) =0,
    \label{alpha2_3=0}
\end{equation}
where $q_{\rm eq}$ is the equilibrium wave number determined by the minimization of the free energy density. At this special point where Eq.~(\ref{alpha2_3=0}) is satisfied, it can be shown that $q_{\rm eq}$ is the same as the wave number obtained by $\partial \alpha_2 / \partial q =0$.

This special point determined by Eq.~(\ref{alpha2_3=0}) is actually the critical point of liquid-solid or disorder-order phase transition, for which the order parameter is the amplitude $A$ of an ordered phase. From Eqs.~(\ref{F_Stripes})--(\ref{F_mbcc}), the one-mode free energy density of any ordered phase is of the same form
\begin{equation}
f = f_u(\bar{n}) + a_2 \alpha_2(q,\bar{n}) A^2 + a_3 \alpha_3(q,\bar{n}) A^3 + a_4 \alpha_4(q) A^4,
\end{equation}
where $f_u$ is given by Eq.~(\ref{fl}), $\alpha_4(q)=E_0+E_{1122}q^4$, and $a_2$, $a_3$, and $a_4$ are constants, with $a_2=1$, $a_3=0$, and $a_4=-1/4$ for stripe or lamellar, $a_2=3$, $a_3=\mp 1$, and $a_4=-15/4$ for triangular or honeycomb, and $a_2=6$, $a_3=\mp 4$, and $a_4=-45/2$ for bcc or mbcc phase. Applying the conditions for the critical point, i.e., $\partial f / \partial A =0$ for the equilibrium state as well as $\partial^2 f / \partial A^2 =0$ and $\partial^3 f / \partial A^3 =0$, it is straightforward to obtain the same condition $\alpha_2=\alpha_3=0$ [i.e., Eq.~(\ref{alpha2_3=0})] for all the phases. The solution of Eq.~(\ref{alpha2_3=0}) leads to Eqs.~(\ref{eps_special_point}) and (\ref{nbar_special_point}), as shown in Appendix \ref{sec:Special_Point}. The same results for the location of this critical point can be obtained beyond the one-mode approximation, as found in our numerical calculations up to three modes.

In addition to the critical point, introducing the $D_{11}$ term also changes the stability and coexistence regimes of different phases in the phase diagram. The phase diagram is symmetric about $\bar{n}=0$ when $D_{11}=0$ [Fig.~\ref{fig:Fig5}(a)] but becomes asymmetric when $D_{11}\neq 0$. For positive $D_{11}$ such as $D_{11}=1$ in Fig.~\ref{fig:Fig5}(b), the phase stability regimes for triangular and bcc structures are reduced while those for their inverse structures (honeycomb and mbcc) enlarged, as compared to the case of $D_{11}=0$. The opposite situation occurs for negative values of $D_{11}$ such as $D_{11}=-1$ shown in Fig.~\ref{fig:Fig5}(c).

\subsection{Elastic constants
\label{sec:elasticity}}

\begin{figure*}
  \centerline{\includegraphics[width=\textwidth]{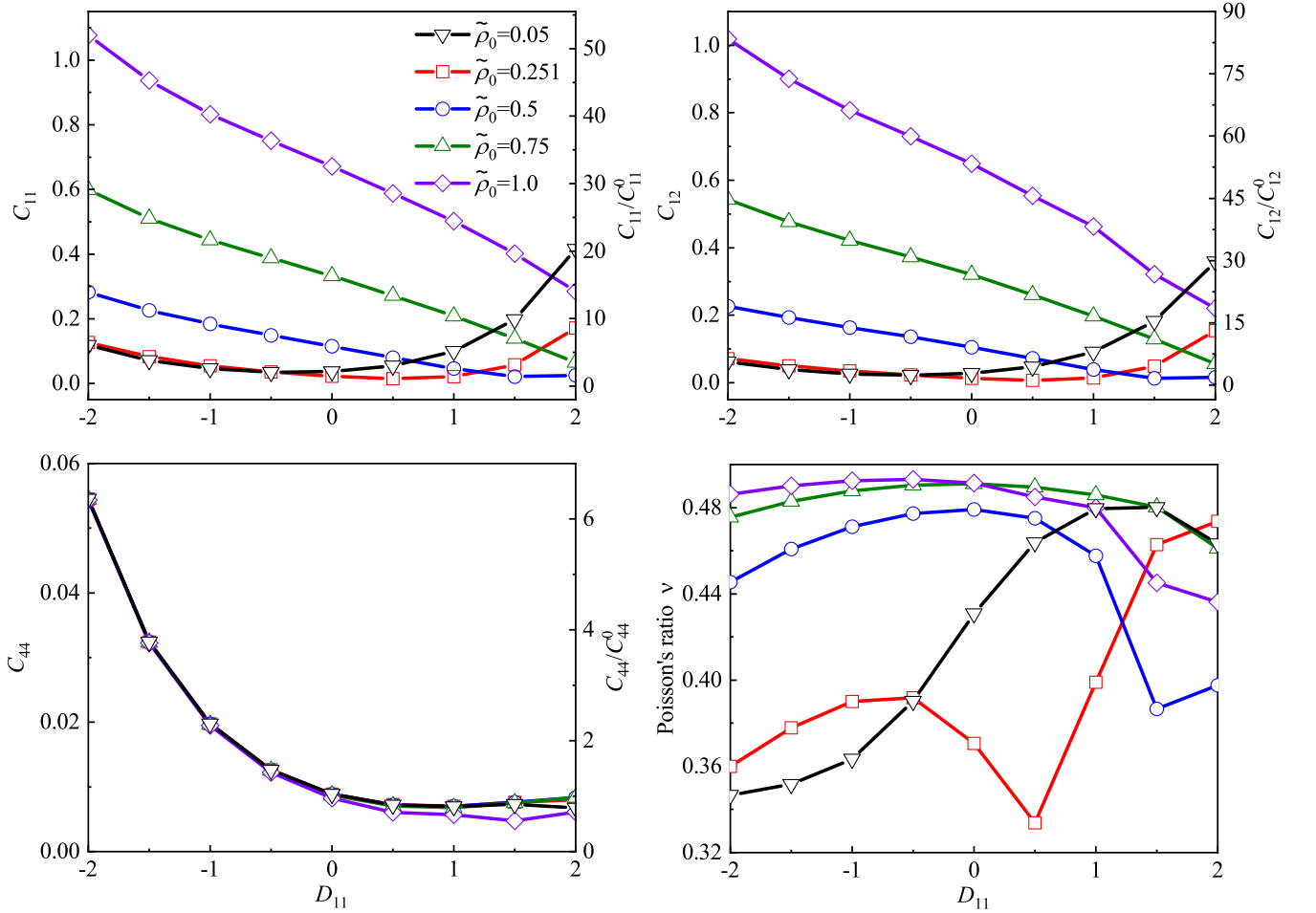}}
  \caption{Elastic constants for bcc phase as a function of $D_{11}$, at $\epsilon=0.1$ (with $C_0=-0.9$) and for five values of $\tilde{\rho}_0=0.05$, $0.251$, $0.5$, $0.75$, and $1$. The other parameters are the same as those of Fig.~\ref{fig:Fig5}. For each $D_{11}$ value, the average density variation in the unstrained state, $\bar{n}_{\rm unstrained}$, is chosen as the mid point of the bcc phase regime in the corresponding phase diagram, i.e., $\bar{n}_{\rm unstrained}=-0.0640$, $-0.0875$, $-0.1122$, $-0.1435$, $-0.1885$, $-0.2518$, $-0.3369$, $-0.4465$, and $-0.5884$ for $D_{11}$ ranging from $-2$ to $+2$ with a step size of $0.5$. Results of elastic constants $C_{ij}$ are also rescaled by $C^0_{ij}$ which are the values of original PFC with $D_{11}=0$, $\bar{n}_{\rm unstrained}=-0.1885$, and $\tilde{\rho}_0=0.251$.}
\label{fig:Fig6}
\end{figure*}

We choose the 3D bcc system as an example to reveal the effect of three-point correlation on the system elastic property. As has been discussed in Ref.~\cite{prb144112}, the isothermal elastic constants of stressed and unstressed systems should be calculated by
\begin{equation}
C_{ijkl}=\frac{1}{V_0}\left. \frac{\partial^2F}{\partial\xi_{ij}\partial\xi_{kl}} \right|_{\xi=0},
\label{Cij}
\end{equation}
where $V_0$ is the volume of the initial undeformed state and $\xi_{ij}$ is an effective finite strain tensor defined by
\begin{equation}
\xi_{ij}=\varepsilon_{ij}-\frac{1}{2}\varepsilon_{ik}\varepsilon_{kj}+\frac{1}{2}\varepsilon_{ij}\varepsilon_{kk},
\label{xi}
\end{equation}
with $\varepsilon_{ij}$ the infinitesimal or linear strain tensor. The strained-state free energy is of the form
\begin{equation}
F_{\rm strained}=\underset{{A}}{\mathrm{min}}\mathcal{F}({A},{\mathbf{q}^{(\rm strained)}},{\bar{n}_{\rm strained}},{V_{\rm strained}}),
\label{F_strained}
\end{equation}
with $A$, $\mathbf{q}$, $\bar{n}$, and volume $V$ all varying with the applied strain. It is worth emphasizing that the values of elastic constants depend on the choice of a non-negative rescaled reference-state density $\tilde{\rho}_{0}$ used in the definition of rescaled PFC density variation field $n$. When the system is strained or deformed, the corresponding variation of the average density variation $\bar{n}$ is governed by the relation \cite{prb144112}
\begin{equation}
\bar{n}_{\rm strained}=\frac{V_{\rm unstrained}}{V_{\rm strained}}(\bar{n}_{\rm unstrained}+\tilde{\rho}_{0})-\tilde{\rho}_{0},
\label{nbar_strained}
\end{equation}
to satisfy the condition of constant total number of particles. The model parameters of original PFC (with $D_{11}=0$) have been matched to bcc Fe, giving $\tilde{\rho}_0=0.251$, and the corresponding elastic constants $C_{ij}^0$ (using Voigt notation) have been identified  \cite{prb144112}.

Here we further calculate the elastic constants of various bcc systems with varying $D_{11}$ and $\tilde{\rho}_0$ which correpond to different types of materials, based on Eqs.~(\ref{Cij})--(\ref{nbar_strained}) and the one-mode free energy density of Eq.~(\ref{F_bcc}). The $\bar{n}_{\rm unstrained}$ for each $D_{11}$ is determined by the mid point of the stable regime of bcc phase in the $\epsilon$-$\bar{n}$ phase diagram at a given temperature parameter $\epsilon$ (e.g., $\epsilon=0.1$ as used in this work). The corresponding results are presented in Fig.~\ref{fig:Fig6}. We have chosen five values of $\tilde{\rho}_0$ to show how it affects the elastic property. Figure \ref{fig:Fig6} indicates that $C_{44}$ has a much weaker dependence on $\tilde{\rho}_0$ as compared to $C_{11}$ and $C_{12}$, while each of them changes significantly across at least a certain range of $D_{11}$. $C_{11}$ and $C_{12}$ exhibit similar behaviors of variation with respect to the change in $\tilde{\rho}_0$ and $D_{11}$. Their values decrease monotonically with the increase of $D_{11}$ for large enough $\tilde{\rho}_0$, while for small $\tilde{\rho}_0$ (e.g., $\tilde{\rho}_0=0.05$ or $0.251$) a minimum of $C_{11}$ or $C_{12}$ can be found when $D_{11}$ varies. In addition, their values increase with $\tilde{\rho}_0$ when $D_{11}<0$.

Variations of the Poisson's ratio $\nu$ are also given in Fig.~\ref{fig:Fig6}, showing a more complicated behavior. When $\tilde{\rho}_0 > 0.251$, $\nu$ reaches a maximum at some small magnitude of $D_{11}$. When $D_{11} < -0.5$ the values of $\nu$ increase with larger $\tilde{\rho}_0$. The changes of $\nu$ are much more irregular in the other ranges of $D_{11}$ or $\tilde{\rho}_0$, including some sharp variations (e.g., at $\tilde{\rho}_0=0.251$ and $0.5$) which could be attributed to different values of average density $\bar{n}_{\rm unstrained}$ chosen at different $D_{11}$ for the stability of bcc phase.

All these results show that in this PFC modeling the elastic property of a system can be varied effectively over a considerable range. This well demonstrates the efficiency and advantage of this model, which incorporates the important effects of three-point correlation, for describing different materials of the same crystalline symmetry but different elastic properties through an extended parameterization. For example, the elastic constants of bcc Fe (with $\tilde{\rho}_0=0.251$) were underestimated in the original PFC model based only on two-point direct correlation \cite{prb144112}. With the incorporation of three-point correlation with nonzero $D_{11}$ (e.g., $=-1$), the values of $C_{11}$, $C_{12}$, $C_{44}$, and $\nu$ would all increase as shown in Fig.~\ref{fig:Fig6}, yielding a better matching to the real material.

\subsection{Vapor-liquid-solid coexistence for 3D bcc}
\label{sec:VLS_bcc}

Recently we used this PFC model to obtain the vapor-liquid-solid transitions and coexistence for 2D triangular phase \cite{prm103802}. Actually, the model based on Eq.~(\ref{Functional}) is generally applicable to both 2D and 3D systems. As discussed in Sec.~\ref{sec:stability analysis}, the vapor-liquid-solid transition sequence can be realized through choosing the values of model parameters with large magnitude of $D_{11}$ [see Fig.~\ref{fig:Fig2}(c)]. In the following we show an example of how to obtain the three-phase transitions and coexistence for 3D bcc. First, we fix the values of $B_0$, $C_0$, $D_0$, and $E_0$ to determine the properties of vapor and liquid phases. We choose $B_0=-1.875$, $C_0=-5.75$, $D_0=-9$, and $E_0=-6$, so that from Eq.~(\ref{ncoexist}) the coexistence densities for vapor and liquid phases are $\bar{n}_{\rm vapor}=-2.5$ and $\bar{n}_{\rm liquid}=-0.5$, respectively. The corresponding free energy densities are $f_{\rm vapor}=f_{\rm liquid}=-0.391$ since $B_0$ has been chosen to make the common tangent line passing the two coexistence densities horizontal. Following the procedure described in Ref.~\cite{prm103802} to identify three-phase coexistence, the coexistence density for solid phase, $\bar{n}_{\rm solid}$, can be chosen flexibly. Here we choose $\bar{n}_{\rm solid}=-0.1$. In equilibrium the three coexisting phases should have equal chemical potential and pressure, which provides two constraints. To obtain the equilibrium free energy density of bcc phase we also need both $\partial f_{\rm bcc}/\partial q=0$ and $\partial f_{\rm bcc}/\partial A=0$. Now we have four restricted relationships with seven undetermined variables $C_2$, $C_4$, $C_6$, $D_{11}$, $E_{1122}$, $q$, and $A$. To suppress the contributions from higher-order modes we set $C_6=32$ \cite{prm103802}. In addition, the equilibrium wave number and amplitude can be preset as some specific values such as $q_{\rm eq}=1$ and $A_{\rm eq}=0.2$ as used here. Then the remaining four parameters $C_2$, $C_4$, $D_{11}$, and $E_{1122}$ can be determined from the four constraint conditions mentioned above. Values of these parameters are first obtained based on the one-mode bcc free energy density Eq.~(\ref{F_bcc}), and then be slightly adjusted to account for the discrepancy between one-mode and full-mode solutions, as identified from the full PFC simulation to obtain the phase coexistence. All the model parameters used in the full PFC numerical calculations are listed in Table~\ref{tab:table_para}.

\begin{table*}
  \caption{\label{tab:table_para}Model parameters used in numerical calculations of vapor-liquid-solid transitions and coexistence for 3D bcc phase, with the triple point temperature at $\Delta T=0$.}
\begin{ruledtabular}
\begin{tabular}{ccccccccc}
  $B_0$  & $C_0$  & $C_2$  & $C_4$  & $C_6$  & $D_0$  & $D_{11}$  & $E_0$  & $E_{1122}$ \\ \hline
  $-4.732-3\Delta T$ & $-5.844-\Delta T$ & $19.337+1.6\Delta T$ & $57.442-0.8\Delta T$ & $32$ & $-9$
  & $-26.632$ & $-6$ & $-17.835$\\
\end{tabular}
\end{ruledtabular}
\end{table*}

\begin{figure*}
  \centerline{\includegraphics[width=0.95\textwidth]{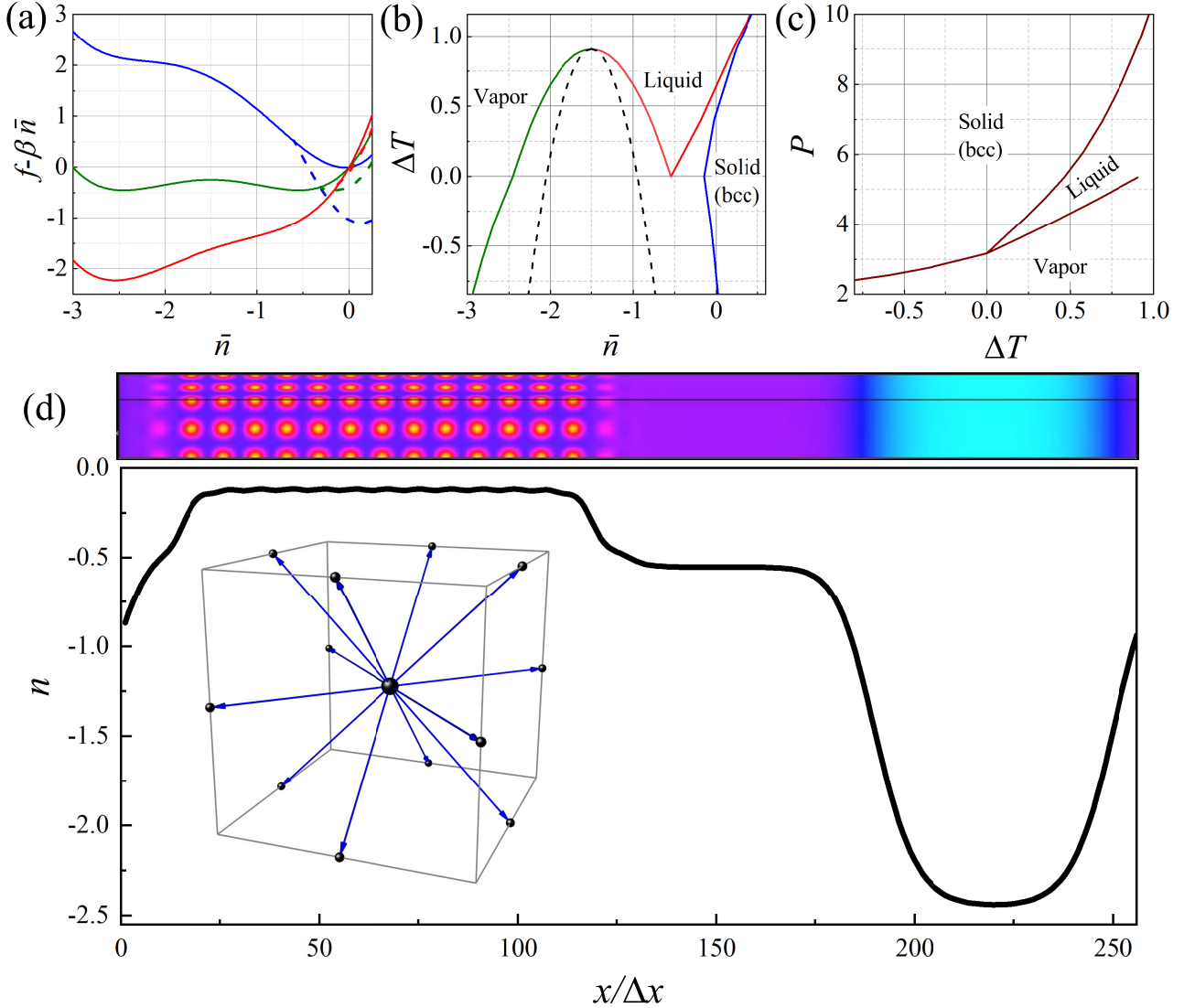}}
  \caption{(a) Free-energy density profiles for uniform (solid curves) and 3D bcc (dashed) phases at temperatures $\Delta T=-0.594$ (blue), $\Delta T=0$ (green), and $\Delta T=0.406$ (red), as calculated from the full PFC Eq.~(\ref{EqPFC}). To give a better illustration, all the curves have been titled by $-\beta \bar{n}$ with $\beta=2.716$. (b) Temperature-density phase diagram, where the black dashed curve corresponds to the vapor-liquid spinodal. (c) The corresponding temperature-pressure phase diagram. (d) The full-PFC dynamical simulation results of three-phase coexistence at $\Delta T=0$. Upper panel: A simulation snapshot showing the solid-liquid-vapor coexistence. Bottom panel: The corresponding density profile along the $x$ direction, where the density has been averaged over the $yz$ cross section; the diffraction pattern is shown as an inset. The parameters used are listed in Table~\ref{tab:table_para}.
  }
  \label{fig:Fig7}
\end{figure*}

The accurate values of equilibrium free energy for bcc phase is obtained through numerically solving the full PFC dynamical Eq.~(\ref{EqPFC}). A single unit cell is used with periodic boundary conditions. To enhance the computational efficiency, we set up the initial conditions by using either the density field $n(\mathbf{r})$ obtained from the one-mode approximation or the existing simulation outcome of equilibrium $n(\mathbf{r})$ profile with close value of average density $\bar{n}$, and evolve the system up to a steady state with negligible time variation of free energy density. For each set of $\bar{n}$ and temperature $\Delta T$, We vary the numerical grid spacings $\Delta x$, $\Delta y$, and $\Delta z$ to find the minimum of the corresponding free energy density and the equilibrium wave number $q_{\rm eq}$. Some examples of the equilibrium free energy density profiles as a function of $\bar{n}$ are shown in Fig.~\ref{fig:Fig7}(a). The common tangent construction is then applied to determine the densities of liquid-solid or vapor-solid coexistence, with the vapor-liquid-solid coexistence realized at $\Delta T=0$. The resulting temperature-density ($\Delta T$ vs $\bar{n}$) phase diagram is given in Fig.~\ref{fig:Fig7}(b).

We also compute the corresponding temperature-pressure ($\Delta T$ vs $P$) phase diagram based on the $f$-$\bar{n}$ curves obtained from the full PFC numerical calculations, as presented in Fig.~\ref{fig:Fig7}(c). Values of the equilibrium pressure $P$ can be calculated from the free energy density $f(\bar{n})$, with details given in Ref.~\cite{prm103802}. Results of both two diagrams are consistent with the known properties of vapor-liquid-solid transitions and coexistence in real materials and previous simulations, such as the temperature dependence of coexistence densities, triple point, and critical point.

To further verify the phase behavior described above, we have conducted a dynamical simulation of 3D vapor-liquid-solid coexistence based on Eq.~(\ref{EqPFC}) using the parameter values listed in Table~\ref{tab:table_para} for bcc phase. The system is set at $\Delta T=0$, and the initial configuration is set up as half of the simulated slab being occupied by a bcc solid with $\bar{n}=-0.14$ and the other half by two equal-volume homogeneous regions with $\bar{n}=-2.25$ and $-0.75$, respectively. During the time evolution the system remains as a mixture of vapor, liquid, and bcc solid phases, as seen in the simulation snapshot and the corresponding diffraction pattern shown in Fig.~\ref{fig:Fig7}(d). The average densities of each phase change from the initial setup values to $\bar{n}=-2.44$, $-0.55$, and $-0.12$ for vapor, liquid, and solid phases, respectively, closely matching the three-phase coexistence densities given in the equilibrium phase diagram of Fig.~\ref{fig:Fig7}(b). An example of the cross-section averaged density profile obtained from our simulation is depicted in Fig.~\ref{fig:Fig7}(d).

The multiphase modeling that is made available in this approach would be important for the study of some material growth and evolution processes, particularly those involving the coexistence and interplay between vapor, liquid, and solid phases. A typical example is the surface-melting mediated crystal growth near the triple point \cite{NatComm239}, which makes use of the property that on solid surface a premelted thin liquid layer would form and separate the underneath growing crystal and outside vapor. The multiphase characteristics, particularly the formation of premelting surface films and the resulting three-phase coexistence, then play a crucial role in determining the growth dynamics and system evolution. These and similar types of multiphase material processes and the underlying mechanisms can be readily modeled and explored through this PFC approach incorporating three-point correlation, which well describes the vapor-liquid-solid coexistence and transition as demonstrated above.

\subsection{Liquid-stripe or liquid-lamellar coexistence}
\label{sec:lid_str_coex}

In all the previous PFC models, the stripe or lamellar phase cannot, except for the one-dimensional case, coexist with the uniform phase, as seen in Fig.~\ref{fig:Fig5}. We will demonstrate in the following that the coexistence between liquid and stripe or lamellar phases can be achieved in either 2D or 3D through the adjustment and control of model parameters particularly with the contributions ($D_{11}$ and $E_{1122}$) from high-order direct correlations. The liquid-hexagonal coexistence in 3D, which is absent in the original PFC, can be also made accessible if following the similar approach.

\begin{figure*}
  \centerline{\includegraphics[width=\textwidth]{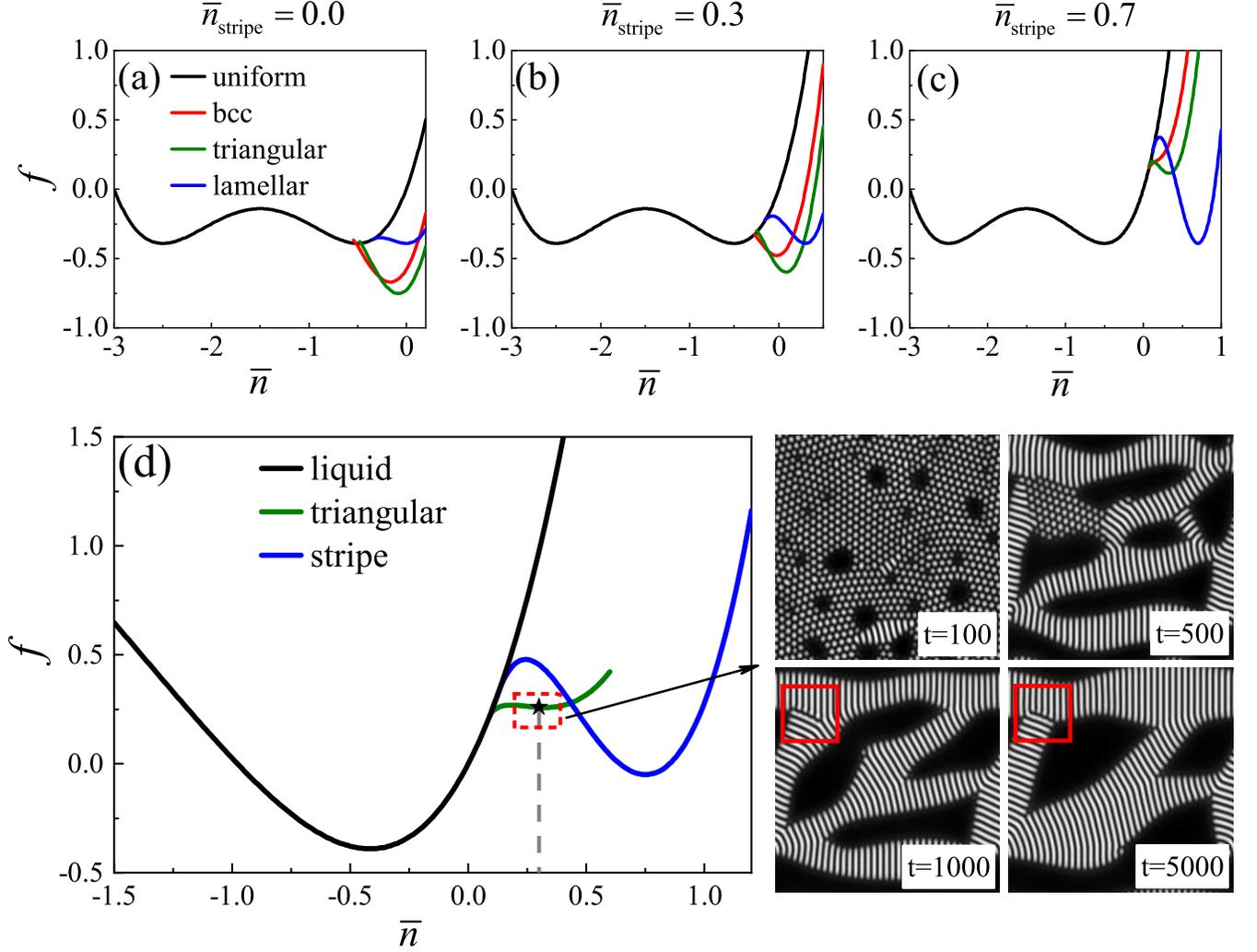}}
  \caption{(a)--(c): Equilibrium free-energy densities for uniform, lamellar, triangular rod, and bcc phases in 3D as a function of $\bar n$ in the one-mode approximation, based on three different values of $\bar{n}_{\rm stripe}$ as indicated. (d) Free energy density profiles for liquid, stripe, and triangular phases in 2D as obtained from numerical solution of the full PFC Eq.~(\ref{EqPFC}). Also shown are four snapshots of the system dynamical evolution starting from a random initial condition of $\bar{n}=0.3$. The red boxed region indicates a high-angle grain boundary of the stripe pattern in the stripe-liquid coexisting state. The parameters used are $B_0=-1.875$, $C_6=32$, $D_0=-9$, $E_0=-6$, and $C_0=-5.75$ for (a)--(c) and $C_0=-6.75$ for (d). In addition, $(C_2,C_4,D_{11},E_{1122})=(15.2,57.8,-30.9,-53.8)$ for (a), $(C_2,C_4,D_{11},E_{1122})=(12.1,56.2,-67.6,-186.1)$ for (b), and $(C_2,C_4,D_{11},E_{1122})=(50.3,72.8,-195.9,-558.5)$ for (c) and (d).}
\label{fig:Fig8}
\end{figure*}

We base the following analysis on the one-mode free energy densities given in Eqs.~(\ref{F_Stripes}), (\ref{F_Rods}), and (\ref{F_bcc}). The key adjustable parameter is the stripe- or lamellar-phase density $\bar{n}_{\rm stripe}$ at the vapor-liquid-ordered phase coexistence. Similar to the procedure described above in Sec.~\ref{sec:VLS_bcc}, we first determine the properties of vapor and liquid phases by setting $B_0=-1.875$, $C_0=-5.75$, $D_0=-9$ and $E_0=-6$. The vapor-liquid coexistence densities can be obtained from Eq.~(\ref{ncoexist}) as $\bar{n}_{\rm vapor}=-2.5$ and $\bar{n}_{\rm liquid}=-0.5$. Next we chose a fixed value of $\bar{n}_{\rm stripe}$ as the density of stripe or lamellar phase when coexisting with liquid and vapor. For simplicity, the equilibrium wave number $q_{\rm eq}$ and the equilibrium amplitude $A_{\rm eq}$ are chosen as 1 and 0.4, respectively, and $C_6=32$ is used as before. The other four parameters, $C_2$, $C_4$, $D_{11}$, and $E_{1122}$, can then be identified for conditions satisfying the three-phase coexistence at each value of $\bar{n}_{\rm stripe}$. Some sample results are given in Figs.~\ref{fig:Fig8}(a)--\ref{fig:Fig8}(c) for $\bar{n}_{\rm stripe}$ varying from 0 to 0.7, showing that the free energy densities of bcc and hexagonal phases increase with larger value of $\bar{n}_{\rm stripe}$ and exceed that of the lamellar phase at large enough $\bar{n}_{\rm stripe}$ [see Fig.~\ref{fig:Fig8}(c)], resulting in the coexistence between lamellar and uniform phases.

To verify these analyses we conduct 2D numerical simulations of the full PFC Eq.~(\ref{EqPFC}) based on the parameters identified from Fig.~\ref{fig:Fig8}(c), with results presented in Fig.~\ref{fig:Fig8}(d). Since here we focus on the liquid-stripe coexistence, the value of $C_0$ is changed from $-5.75$ to $-6.75$ to destabilize the vapor phase, while all the other parameters remain unchanged. The equilibrium free energy density profiles for liquid, stripe, and triangular phases are first calculated through the full-mode numerical solution of the PFC Eq.~(\ref{EqPFC}) in 2D, as shown in the left panel of Fig.~\ref{fig:Fig8}(d). We then choose $\bar{n}=0.3$, which locates within the liquid-stripe coexistence regime [as marked on the $f$-$\bar{n}$ curve in Fig.~\ref{fig:Fig8}(d)], to run a dynamical simulation test. The initial configuration is set up as a homogeneous state with random initial condition of $\bar{n}=0.3$. The subsequent evolution dynamics is governed by Eq.~(\ref{EqPFC}). Four simulation snapshots are shown in the right panels of Fig.~\ref{fig:Fig8}(d), corresponding to time $t=100$, $500$, $1000$, and $5000$, respectively. A transient of liquid-triangular phase coexistence is found at the early stage ($t<100$), after which the stripe phase starts to form and grows gradually to replace the triangular phase, resulting in the coexistence of liquid, stripe, and triangular structures (see e.g., $t=500$). At the late time stage ($t \geq 1000$) only the coexistence between liquid and stripe phases is observed. The evolution of the stripe pattern is accompanied by the dynamics of topological defects including disclinations, dislocations, and grain boundaries, consistent with the known pattern-forming and evolution process for the stripe or smectic phase. An example of a time-evolving high-angle grain boundary is shown in the panels of late time stage in Fig.~\ref{fig:Fig8}(d) ($t=1000$ and $5000$, marked with red boxes). Furthermore, as enabled by our PFC modeling with the incorporation of three-point correlation, the results here demonstrate the additional dynamical feature of isotropic-smectic coexistence and evolution which is absent in previous PFC-type models but important in the study of ordering process of some soft material systems such as colloidal rods. These results of dynamic evolution further evidence the efficacy of three-point correlation on controlling phase behaviors in the PFC modeling.

\section{Summary}
\label{sec:summary}

We have studied the effects of three-point direct correlation on the properties of phase stability, coexistence, and system elastic response in the PFC modeling. Compared to the original PFC model containing only the contribution from two-point direct correlation, the stability regimes of ordered phases are significantly affected by the incorporation of the three-point correlation. From the instability analysis on the supercooled or supersaturated uniform state, much richer and broader conditions for the stability of ordered phase can be identified with the effect of the $D_{11}$ term, a contribution from the three-point correlation. In the $\epsilon$-$\bar{n}$ phase diagrams the stability regimes of bcc and triangular phases expand (or shrink) with negative (or positive) value of $D_{11}$, while those of their inverse counterparts (i.e., mbcc and honeycomb) behave in the opposite way, leading to asymmetric phase diagrams.  Of particular interest is the critical point in these phase diagrams, exhibiting as the merging point of all the phase boundaries, which becomes adjustable with varying nonzero $D_{11}$ values. In addition, the introduction of three-point correlation enables the modeling of different materials of the same crystalline structure but different elastic properties, through a viable control of system elastic constants that can better match various real materials.

Our results also show that the vapor-liquid-solid transitions and coexistence, which are important in the multiphase material growth and evolution processes, can be well realized in our PFC model incorporating effects of multi-point direct correlations, as demonstrated for 3D bcc phase in this work. The liquid-stripe or liquid-lamellar coexistence, which is not accessible in the original or other PFC models but occurs in soft-matter ordering systems (e.g., isotropic-smectic coexistence and transition in colloidal liquid crystals or rods), can be achieved as well through parameter control. 3D and 2D PFC dynamical simulations have been conducted to verify these findings and further prove the effects of three-point direct correlation.

In short, our study has demonstrated that contributions from three-point direct correlation can be used to control and adjust the phase stability regimes, the relative stability of different ordered phases, phase coexistence, elastic properties, and the transition sequence among vapor, liquid, and ordered phases in either solid or soft matter systems. This ability of phase and elastic-property control makes our PFC approach a valuable tool for modeling a wide variety of real material systems.

\begin{acknowledgments}
Z.-L.W.~acknowledges support from the China Postdoctoral Science Foundation (Grant No.~2020M670275).
Z.R.L.~acknowledges support from the National Natural Science Foundation of China (Grant No.~21773002).
W.D.~acknowledges support from the National Natural Science Foundation of China (Grant No.~51788104), the Ministry of Science and Technology of China (Grant No.~2016YFA0301001), and the Beijing Advanced Innovation Center for Materials Genome Engineering.
Z.-F.H.~acknowledges support from the U.S. National Science Foundation under Grant No. DMR-2006446.
\end{acknowledgments}

\appendix

\section{Linear instability analysis}
\label{sec:Stable_region}

As stated in Sec.~\ref{sec:stability analysis}, the uniform or homogeneous phase is linearly unstable with respect to ordered phase when $\alpha_2 (\bar{n})<0$. Also, to enable the ordered phase $q^2>0$ is required. Hence $\alpha_2 (\bar{n})<0$ and $q^2>0$ are the two necessary but not sufficient conditions for stabilizing the ordered phases. For simplicity, $C_4<0$, $E_0<0$, and $C_6=E_{1122}=0$ are set in the following analysis which can be classified into two cases of $E_0-2D_{11}^2/(9C_4)<0$ and $>0$ respectively.

(i) Based on the formula of $\alpha_2(\bar{n})$ given in Eq.~(\ref{Linear_no_q}), when $E_0-2D_{11}^2/(9C_4)<0$ the $\alpha_2$ vs $\bar{n}$ curve is a convex parabola. According to Eq.~(\ref{nsupercool}), if $b_s^2-4a_sc_s>0$ or equivalently $C_0>C_2^2/(4C_4)+b_s^2/(2a_s)$, $\alpha_2(\bar{n})=0$ has two nonzero real solutions $\bar{n}_+$ and $\bar{n}_-$ (noting that $\bar{n}_->\bar{n}_+$ here). Thus, $\alpha_2(\bar{n})<0$ occurs for $\bar{n}$ in the interval of $(\bar{n}_+, \bar{n}_-)$. The other condition of $q^2>0$ corresponds to the following two situations.

First, suppose $q^2>0$ is satisfied for all the $\bar{n}$ values in the $(\bar{n}_+, \bar{n}_-)$ interval. Obviously, if $D_{11}=0$ this is always true since $q^2$ is then equal to $C_2/2C_4$ (when $E_{1122}=0$) with both $C_2$ and $C_4$ being negative. As also seen in Eq.~(\ref{Linear_q2}), if $D_{11}\neq 0$, $q^2$ is dependent on $\bar{n}$. To satisfy the condition of $q^2>0$ in the whole range of $\bar{n}_+ < \bar{n} < \bar{n}_-$, we need
\begin{eqnarray}
&&(C_2+\frac{2}{3}D_{11}\bar{n}_-)(C_2+\frac{2}{3}D_{11}\bar{n}_+) \nonumber\\
&&=C_2^2-\frac{4(3D_0C_2D_{11}-2C_0D_{11}^2-\frac{C_2^2D_{11}^2}{2C_4})}{9(E_0-\frac{2D_{11}^2}{9C_4})}>0, \label{eq:n+->0}
\end{eqnarray}
leading to
\begin{equation}
\frac{C_2^2}{4C_4}+\frac{b_s^2}{2a_s}<C_0<\frac{3C_2D_0}{2D_{11}}-\frac{9C_2^2E_0}{8D_{11}^2}, \label{eq:C01C02}
\end{equation}
and if defining the midpoint of $\bar{n}_+$ and $\bar{n}_-$ as $\bar{n}_m$,
\begin{equation}
C_2+\frac{2}{3}D_{11}\bar{n}_m=C_2-D_{11}\frac{2(D_0-\frac{C_2D_{11}}{3C_4})}{3(E_0-\frac{2D_{11}^2}{9C_4})}<0 {\rm ~or}~ >0,
\label{eq:C2}
\end{equation}
which yields
\begin{eqnarray}
&& C_2<\frac{2D_0D_{11}}{3E_0}, {\rm ~if~} -\frac{\sqrt{9E_0C_4}}{2}<D_{11}<0, \label{eq:C2_1}\\
&& C_2>\frac{2D_0D_{11}}{3E_0}, {\rm ~if~} 0<D_{11}<\frac{\sqrt{9E_0C_4}}{2}, \label{eq:C2_2}
\end{eqnarray}
respectively. Thus, Eqs.~(\ref{eq:C01C02}), (\ref{eq:C2_1}), and (\ref{eq:C2_2}) need to be followed for the ordered-phase stability at values of $\bar{n}$ within the range of $(\bar{n}_+, \bar{n}_-)$ in this case (i).

Second, if $q^2>0$ is satisfied only for some $\bar{n}$ values in the $(\bar{n}_+, \bar{n}_-)$ interval, we have
\begin{equation}
 (C_2+\frac{2}{3}D_{11}\bar{n}_-)(C_2+\frac{2}{3}D_{11}\bar{n}_+)<0,
 \label{eq:n+-<0}
\end{equation}
instead of Eq.~(\ref{eq:n+->0}), giving
\begin{equation}
C_0>\frac{3C_2D_0}{2D_{11}}-\frac{9C_2^2E_0}{8D_{11}^2}.
\label{eq:C02}
\end{equation}
Since here $E_0-2D_{11}^2/(9C_4)<0$, when $-\sqrt{9E_0C_4}/2<D_{11}<0$, $q^2$ increases monotonically with $\bar{n}$ as seen from Eq.~(\ref{Linear_q2}). Thus the interval of $\bar{n}$ for $q^2>0$ is $(\bar{n}_0, \bar{n}_-)$, where $\bar{n}_0$ is the solution of $q^2=0$. On the other hand, when $0<D_{11}<\sqrt{9E_0C_4}/2$, $q^2$ decreases monotonically with the increase of $\bar{n}$, and hence the $\bar{n}$ interval for $q^2>0$ is $(\bar{n}_+, \bar{n}_0)$. In both situations Eq.~(\ref{eq:C02}) should be obeyed as well to obtain the stability regime of ordered phase.
%In brief, when $C_0>3C_2D_0/(2D_{11})-9C_2^2E_0/(8D_{11}^2)$, the stability regime of ordered phase is $(\bar{n}_0, \bar{n}_-)$ with $-\sqrt{9E_0C_4}/2<D_{11}<0$ or $(\bar{n}_+, \bar{n}_0)$ with $0<D_{11}<\sqrt{9E_0C_4}/2$.

(ii) When $E_0-2D_{11}^2/(9C_4)>0$, $\alpha(\bar{n})$ is a concave parabola as a function of $\bar{n}$. Similar to case (i), when $C_0<C_2^2/(4C_4)+b_s^2/(2a_s)$ so that $b_s^2-4a_sc_s>0$ in Eq.~(\ref{nsupercool}), there are two nonzero real solutions of $\alpha(\bar{n})=0$, i.e., $\bar{n}_+$ and $\bar{n}_-$, but now $\bar{n}_-<\bar{n}_+$. $\alpha(\bar{n})<0$ occurs for $\bar{n}$ lying in the interval of $(-\infty, \bar{n}_-)\cup(\bar{n}_+, +\infty)$.

If $\bar{n}_0<\bar{n}_-$, since in this case (ii) $E_0-2D_{11}^2/(9C_4)>0$, $q^2$ increases monotonically with $\bar{n}$ when $D_{11}<-\sqrt{9E_0C_4}/2$. The $\bar{n}$ interval that satisfies both $\alpha(\bar{n})<0$ and $q^2>0$ for the stability of ordered phase is then $(\bar{n}_0, \bar{n}_-)\cup(\bar{n}_+, +\infty)$. When $D_{11}>\sqrt{9E_0C_4}/2$, $q^2$ decreases monotonically with the increase of $\bar{n}$, and the corresponding stability interval is $(-\infty, \bar{n}_0)$. For both scenarios Eqs.~(\ref{eq:n+->0}) and (\ref{eq:C2}) should be also satisfied, but resulting in different conditions of
\begin{equation}
    \frac{3C_2D_0}{2D_{11}}-\frac{9C_2^2E_0}{8D_{11}^2} < C_0 < \frac{C_2^2}{4C_4}+\frac{b_s^2}{2a_s},
    \label{eq:C02C01}
\end{equation}
and
\begin{eqnarray}
&& C_2>\frac{2D_0D_{11}}{3E_0}, {\rm ~if~} D_{11}<-\frac{\sqrt{9E_0C_4}}{2}, \label{eq:C2_3}\\
&& C_2<\frac{2D_0D_{11}}{3E_0}, {\rm ~if~} D_{11}>\frac{\sqrt{9E_0C_4}}{2}. \label{eq:C2_4}
\end{eqnarray}
To summarize, the stability regime of ordered phase is within the $\bar{n}$ range of $(\bar{n}_0, \bar{n}_-)\cup(\bar{n}_+, +\infty)$ under the conditions of Eqs.~(\ref{eq:C02C01}) and (\ref{eq:C2_3}), or the range of $(-\infty, \bar{n}_0)$ with the conditions of Eqs.~(\ref{eq:C02C01}) and (\ref{eq:C2_4}).

If $\bar{n}_-<\bar{n}_0<\bar{n}_+$, $\alpha(\bar{n})<0$ and $q^2>0$ can be satisfied for $\bar{n}$ within $(\bar{n}_+, +\infty)$ when $D_{11}<-\sqrt{9E_0C_4}/2$, or in the interval of $(-\infty, \bar{n}_-)$ when $D_{11}>\sqrt{9E_0C_4}/2$. In either case Eq.~(\ref{eq:n+-<0}) is also obeyed, giving
\begin{equation}
    C_0<\frac{3C_2D_0}{2D_{11}}-\frac{9C_2^2E_0}{8D_{11}^2}.
\label{Ap_A_Eq4}
\end{equation}
%In summary, when $C_0<3C_2D_0/(2D_{11})-9C_2^2E_0/(8D_{11}^2)$, the stability regime of ordered phase is $(\bar{n}_+, +\infty)$ with $D_{11}<-\sqrt{9E_0C_4}/2$ or $(-\infty, \bar{n}_-)$ with $D_{11}>\sqrt{9E_0C_4}/2$.

If $\bar{n}_0>\bar{n}_+$, to satisfy the conditions of $\alpha(\bar{n})<0$ and $q^2>0$ for the stability of ordered phase, we can obtain the same Eq.~(\ref{eq:C02C01}) from solving Eq.~(\ref{eq:n+->0}), but for different allowed ranges of $\bar{n}$ when following the similar derivation steps. The solution of Eq.~(\ref{eq:C2}) yields
\begin{eqnarray}
&& C_2<\frac{2D_0D_{11}}{3E_0}, {\rm ~if~} D_{11}<-\frac{\sqrt{9E_0C_4}}{2}, \label{eq:C2_5}\\
&& C_2>\frac{2D_0D_{11}}{3E_0}, {\rm ~if~} D_{11}>\frac{\sqrt{9E_0C_4}}{2}. \label{eq:C2_6}
\end{eqnarray}
Eqs.~(\ref{eq:C02C01}) and (\ref{eq:C2_5}) give the ordered-phase stability conditions for the $\bar{n}$ range within $(\bar{n}_0, +\infty)$, while Eqs.~(\ref{eq:C02C01}) and (\ref{eq:C2_6}) correspond to the conditions for the range of $(-\infty, \bar{n}_-)\cup(\bar{n}_+, \bar{n}_0)$.

Finally, if $b_s^2-4a_sc_s<0$ in Eq.~(\ref{nsupercool}), or equivalently $C_0>C_2^2/(4C_4)+b_s^2/(2a_s)$, no real solution exists for $\alpha(\bar{n})=0$ and we have $\alpha(\bar{n})<0$ for all the $\bar{n}$ values when $E_0-2D_{11}^2/(9C_4)>0$ in this case (ii). When $D_{11}<-\sqrt{9E_0C_4}/2$, $q^2$ increases monotonically with the increase of $\bar{n}$ and $q^2>0$ is satisfied in the $\bar{n}$ range of $(\bar{n}_0, +\infty)$. When $D_{11}>\sqrt{9E_0C_4}/2$, $q^2$ decreases monotonically with the increase of $\bar{n}$ and $q^2>0$ is satisfied in the interval of $(-\infty, \bar{n}_0)$. These analyses suggest that when $C_0>C_2^2/(4C_4)+b_s^2/(2a_s)$, the stability regime of ordered phase is $(\bar{n}_0, +\infty)$ for $D_{11}<-\sqrt{9E_0C_4}/2$ or $(-\infty, \bar{n}_0)$ for $D_{11}>\sqrt{9E_0C_4}/2$.

All the above results of linear stability regimes for ordered phase and the corresponding parameter conditions are summarized in Table~\ref{tab:stab_reg}.

\section{Derivation of the critical point for the convergence of all the phase boundaries}
\label{sec:Special_Point}

As described in Sec.~\ref{sec:Phase_diagram}, in the one-mode approximation Eq.~(\ref{alpha2_3=0}) can be applied to identify the location of the critical point in the phase diagram where all the ordered and disorder phases coexist. Using Eqs.~(\ref{F_bcc}) and (\ref{F_mbcc}) for bcc and mbcc phases, we get
\begin{equation}
C_0+D_0\bar{n}+\frac{1}{2}E_0\bar{n}^2-\left(C_2+\frac{2}{3}D_{11}\bar{n}\right)q_{\rm eq}^2+C_4q_{\rm eq}^4=0, \label{A2_A3_0}
\end{equation}
\begin{equation}
\left ( D_0-D_{11}q_{\rm eq}^2 \right ) + E_0\bar{n}=0,
\label{f2_f3_eq_0}
\end{equation}
from $\alpha_2=0$ and $\alpha_3=0$ respectively, where we have assumed $E_{1122}=0$ for simplicity, and $q_{\rm eq}$ is the equilibrium wave number of the bcc or mbcc phase obtained by solving $\partial f_{\rm bcc}/\partial q=0$ and $\partial f_{\rm bcc}/\partial A=0$. This leads to
\begin{equation}
    q_{\rm eq}^2=\frac{b_0+\sqrt{b_0^2-3a_0c_0}}{2a_0},
    \label{bcc_qeq}
\end{equation}
where
\begin{eqnarray}
&& a_0 = C_4 \left ( 405C_4E_0 - 48D_{11}^2 \right ), \nonumber\\
&& b_0 = -72C_4D_0D_{11}-12C_2D_{11}^2+405C_2C_4E_0 \nonumber\\
&& \qquad\, +2D_{11}\bar{n}(99C_4E_0-4D_{11}^2), \nonumber\\
&& c_0 = -48C_2D_0D_{11}+32C_0D_{11}^2+135C_2^2E_0 \nonumber\\
&& \qquad\, +132C_2D_{11}E_0\bar{n}+44D_{11}^2E_0\bar{n}^2. \nonumber
\end{eqnarray}
On the other hand, Eq.~(\ref{f2_f3_eq_0}) gives
\begin{equation}
q_{\rm eq}^2=\frac{\bar{n}E_0+D_0}{D_{11}}.
\label{f3_qeq}
\end{equation}
The two solutions of $q_{\rm eq}^2$ in Eqs.~(\ref{bcc_qeq}) and (\ref{f3_qeq}) should be equivalent to each other. From this requirement and Eq.~(\ref{A2_A3_0}) we can then derive the expression of $\bar{n}$. For the parameter values used for the phase diagrams of Fig.~\ref{fig:Fig5}, i.e., $D_0=0$, $C_2=-2$, $C_4=-1$, and $E_0=-6$, we have
\begin{widetext}
\begin{equation}
\bar{n}_{\rm bcc}=\frac{-2340D_{11}+111D_{11}^3 \mp 2\sqrt{14580D_{11}^4-1071D_{11}^6+C_0D_{11}^4(14580-1476D_{11}^2+41D_{11}^4)}}{14580-1476D_{11}^2+41D_{11}^4},
\label{bcc_nbar}
\end{equation}
\end{widetext}
with ``$-$'' for $D_{11}>0$ and ``$+$'' for $D_{11} \leq 0$. Following the similar process, we can also obtain the result of $\bar{n}_{\rm hex}$ for the hexagonal phase (i.e., triangular or honeycomb rods) satisfying the condition that the coefficients of $A^2$ and $A^3$ terms equal zero simultaneously. From Eq.~(\ref{F_Rods}) or (\ref{F_graphene}) we get
\begin{widetext}
\begin{equation}
\bar{n}_{\rm hex}=\frac{-810D_{11}+39D_{11}^3 \mp \sqrt{4860D_{11}^4-369D_{11}^6+C_0D_{11}^4(4860-504D_{11}^2+14D_{11}^4)}}{4860-504D_{11}^2+14D_{11}^4}.
\label{hex_nbar}
\end{equation}
\end{widetext}
By solving $\bar{n}_{\rm bcc}=\bar{n}_{\rm hex}=\bar{n}_{\rm supercool}$ [with $\bar{n}_{\rm supercool}$ determined by Eq.~(\ref{nsupercool})], we can obtain the analytic results of $\epsilon$ (noting $\epsilon=1+C_0$) and $\bar{n}$ for the critical point, i.e., Eqs.~(\ref{eps_special_point}) and (\ref{nbar_special_point}).

\bibliography{pfc_3point_correlation_references}

\end{document}